\documentclass[aps,prd,twocolumn,nofootinbib,showpacs,superscriptaddress,floatfix]{revtex4-1}

\usepackage{graphicx,enumerate}
\usepackage{amsmath,epsfig}

\newcommand{\be}{\begin{equation}}
\newcommand{\ee}{\end{equation}}
\newcommand{\ba}{\begin{eqnarray}}
\newcommand{\ea}{\end{eqnarray}}
\newcommand{\nn}{\nonumber}
\newcommand{\MGBK}{{\mbox{\tiny MGBK}}}
\newcommand{\I}{{\cal{I}}}

\begin{document}

\title{Regular black hole metric with three constants of motion}

\author{Tim Johannsen}
\affiliation{Department of Physics and Astronomy, University of Waterloo, Waterloo, Ontario, N2L 3G1, Canada}
\affiliation{Canadian Institute for Theoretical Astrophysics, University of Toronto, Toronto, Ontario M5S 3H8, Canada}
\affiliation{Perimeter Institute for Theoretical Physics, Waterloo, Ontario, N2L 2Y5, Canada}

\date{\today}

\begin{abstract}

According to the no-hair theorem, astrophysical black holes are uniquely characterized by their masses and spins and are described by the Kerr metric. Several parametric spacetimes which deviate from the Kerr metric have been proposed in order to test this theorem with observations of black holes in both the electromagnetic and gravitational-wave spectra. Such metrics often contain naked singularities or closed timelike curves in the vicinity of the compact objects that can limit the applicability of the metrics to compact objects that do not spin rapidly, and generally admit only two constants of motion. The existence of a third constant, however, can facilitate the calculation of observables, because the equations of motion can be written in first-order form. In this paper, I design a Kerr-like black hole metric which is regular everywhere outside of the event horizon, possesses three independent constants of motion, and depends nonlinearly on four free functions that parameterize potential deviations from the Kerr metric. This metric is generally not a solution to the field equations of any particular gravity theory, but can be mapped to known four-dimensional black hole solutions of modified theories of gravity for suitable choices of the deviation functions. I derive expressions for the energy, angular momentum, and epicyclic frequencies of a particle on a circular equatorial orbit around the black hole and compute the location of the innermost stable circular orbit. In addition, I write the metric in a Kerr-Schild-like form, which allows for a straightforward implementation of fully relativistic magnetohydrodynamic simulations of accretion flows in this metric. The properties of this metric make it a well-suited spacetime for strong-field tests of the no-hair theorem in the electromagnetic spectrum with black holes of arbitrary spin.

\end{abstract}


\maketitle


\section{Introduction}

According to the no-hair theorem, isolated and stationary black holes in general relativity are uniquely characterized by their masses $M$ and spins $J$ and are described by the Kerr metric. This metric is the unique stationary, axisymmetric, asymptotically flat, vacuum solution of the Einstein field equations which contains an event horizon but no closed timelike curves in the exterior domain \cite{NHT,rigidity}. Thanks to the no-hair theorem, all astrophysical black holes are expected to be Kerr black holes. While observational evidence suggests the existence of event horizons in astrophysical black holes (see the discussion in, e.g., Ref.~\cite{Narayan}), a proof of the validity of the no-hair theorem is still lacking.

Astrophysical black holes, however, will not be perfectly stationary or exist in perfect vacuum because of the presence of other objects or fields such as stars, accretion disks, or dark matter, which could alter the Kerr nature of the black hole. Nonetheless, under the assumption that such perturbations are so small to be practically unobservable, one can argue that astrophysical black holes are indeed described by the Kerr metric. This is the assumption I make in this paper. This is usually a good approximation for supermassive black holes in the centers of galaxies and for stellar-mass black holes in X-ray binaries, which are typically separated from their respective companion stars by $\sim1~{\rm AU}$ (see, e.g., Ref.~\cite{McCR06}).

Several model-independent strong-field tests of the no-hair theorem have been suggested using gravitational-wave observations of extreme mass-ratio inspirals (EMRIs)~\cite{Ryan95,EMRI,kludge,CH04,GB06,Brink08,Gair08,Apostolatos09,VH10,VYS11,GY11} and electromagnetic observations of accretion flows~\cite{PaperI,PaperII,PaperIII,EM,BambiDiskJet,BambiBarausse11,PJ11,PaperIV,BambiQPO,BambiIron,BambiOthers,Krawcz12,OSCO}. These tests are designed in a phenomenological approach that encompasses large classes of modified theories of gravity instead of focusing on any particular gravity theory. In this approach, the underlying theory is usually unknown, but it is assumed that particles in this theory move along geodesics. The goal is, then, to gain insight in this theory through observations. See Ref.~\cite{reviews} for reviews of such tests.

Other tests of the no-hair theorem exist that include the observation of gravitational ringdown radiation of perturbed black holes after a merger with another object, which tests whether the end-state of the merger is a Kerr black hole \cite{ringdown}, as well as several weak-field tests in the electromagnetic spectrum such as those obtained from the monitoring of close stellar orbits around Sgr~A*~\cite{Sgr} and pulsar/black-hole binaries~\cite{WK99}.

While it is sufficient for tests in the weak-field regime to rely on a parameterized post-Newtonian approach (PPN; e.g.,~\cite{Will93}), the model-independent strong-field tests described above require a modified spacetime which deviates from the Kerr metric in parametric form. Several such parametric frameworks have been created, within which possible observational signatures of a Kerr-like black hole can be explored (e.g.,~\cite{MN92,CH04,GB06,VH10,VYS11,JPmetric}). These metrics can deviate slightly to severely from the Kerr metric, and observables can be studied in terms of one or more free parameters. All of these metrics reduce to the Kerr metric if the deviations vanish. Since general relativity has to date only been marginally tested in the strong-field regime \cite{PsaltisLRR}, deviations from the Kerr metric could be either small or large as long as they are consistent with current weak-field tests (see \cite{WillLRR}).

If a deviation from the Kerr metric is detected, there are two possible interpretations. Within general relativity, the object in this case cannot be a black hole, but is instead a stable stellar configuration or else an exotic object~\cite{CH04,Hughes06}. However, if the compact object is known to possess an event horizon, then the no-hair theorem is falsified and four-dimensional general relativity is only approximately valid in the strong-field regime. 

Because of the no-hair theorem, all parametrically deformed Kerr spacetimes have to violate at least one of the assumptions of this theorem. These metrics are generally stationary, axisymmetric, and asymptotically flat. If such a metric is also Ricci-flat it either does not harbor a black hole or contains singularities or regions with closed timelike curves outside of the event horizon. Such pathologies may also exist, if the metric is not Ricci-flat. The presence of pathologies in a given metric hampers using the metric for strong-field tests of the no-hair theorem in both the gravitational-wave and electromagnetic spectra, because physical processes in the immediate vicinity of the black hole at radii comparable to the innermost-stable circular orbit (ISCO) cannot be modeled properly. Future gravitational-wave detectors, however, will be most sensitive to EMRIs that occur at radii roughly in the range between the innermost stable orbit and $10-20M$ and accretion disks are often modeled to terminate at the ISCO.

For certain applications such as ray-tracing simulations of the electromagnetic radiation emitted from geometrically thin accretion flows a cutoff radius can be imposed which acts as an artificial horizon and shields the adverse effects of the pathological regions from distant observers. While, in principle, information on an enclosed pathology will be encoded in the physical boundary conditions of the cutoff which could still affect the causal past of particles or observers outside of the cutoff, in practice, any photon that reaches the cutoff radius can be excluded from the simulation without altering the observed signal (see the discussion in, e.g., Ref.~\cite{PaperIV}). For rapidly-spinning black holes, however, the ISCO can be arbitrarily close to the event horizon. Therefore, such a metric can only be used for strong-field tests with at most moderately-spinning black holes as long as the ISCO lies outside of the cutoff radius. See Ref.~\cite{Joh13} for further discussion.

Nonetheless, it is possible to design Kerr-like metrics that are free of such pathologies outside of the central object \cite{JPmetric,VYS11}. These two metrics parameterize deviations from the Kerr metric in generic form. While the metric proposed in Ref.~\cite{VYS11} generally describes a black hole, the metric of Ref.~\cite{JPmetric} harbors a black hole only for small perturbations away from the Kerr metric and generally describes a naked singularity. In this case, however, it is always possible to choose a cutoff radius between the surface of the naked singularity and the ISCO, which still allows for the study of the electromagnetic emission from, e.g., geometrically thin accretion disks in this metric \cite{Joh13}. The metric of Ref.~\cite{VYS11} also admits an approximately (i.e., for small deviations from the Kerr metric) conserved third constant of motion in addition to the two associated with stationarity and axisymmetry, which is useful for the construction of approximate EMRI waveforms \cite{GY11}. These metrics, therefore, can be used for strong-field tests of the no-hair theorem even for large values of the spin \cite{Joh13}. 

The structure of these metrics, however, has several disadvantages. The metric of Ref.~\cite{JPmetric} depends only on one set of deviation parameters. In general relativity, stationary, axisymmetric, asymptotically flat, vacuum metrics can be written in terms of four independent functions \cite{WaldBook}, and one would expect that such metrics in more general theories of gravity should depend on at least four independent functions and, thus, on four sets of deviation parameters. Likewise, it not obvious how to implement three-dimensional fully relativistic hydrodynamic simulations of accretion flows in the presence of a cutoff \cite{Joh13}. The metric of Ref.~\cite{VYS11} depends only linearly on four sets of deviation parameters. This may not be the optimal form to parameterize such deviations, because there is no reason to believe that they are small.

Such a dependence on the deviation parameters is sufficient for tests of the no-hair theorem with EMRI observations as long as the deviations are assumed to be small, because an extension to include large deviations would require knowledge of the strong-field radiative dynamics in the underlying theory (e.g., wave emission and polarization, radiation reaction, matter coupling). Strong-field tests of the no-hair theorem in the electromagnetic spectrum, however, are not a priori limited to the study of small deviations from the Kerr metric, because these are performed in a stationary black hole spacetime, where the metric serves as a fixed background. Therefore, the dynamical properties of the gravity theory are not important and the underlying field equations do not have to be known.

Electromagnetic tests of the no-hair theorem, then, only require a suitable accretion flow model, for which additional assumptions regarding the coupling of matter to electromagnetic fields in the modified theory have to be made. The simplest approach is to assume that all electromagnetic interactions are governed by the laws of Maxwell electrodynamics. This is a reasonable choice, given that, at present, there is no evidence for violations of this theory at the classical level (see, e.g., \cite{WillLRR}). In many geometrically thin accretion disk models, the disk plasma is constrained to move along circular equatorial orbits and photons are emitted by plasma particles in their respective local rest frames according to the prescriptions of elementary atomic physics without any subsequent interactions. The resulting observed spectra can be modeled using a variety of different ray-tracing algorithms (e.g., \cite{RTA}).

In fully relativistic magnetohydrodynamic simulations of accretion flows, the dynamics of the flow plasma and electromagnetic fields are governed by the equations expressing the conservation of particle number and the stress-energy of the particles. Such simulations have, so far, only been performed in general relativity (e.g., \cite{GMcKT03,McKG04}), where the flow particles are usually described as a perfect fluid, but these can likewise be carried out in a suitable Kerr-like metric. In many numerical calculations involving either geometrically thin or thick accretion flows, however, a linear dependence on the deviation parameters cannot be enforced, because the equations that govern the evolution of particles and fields such as the geodesic equations are usually nonlinear. It is, therefore, desirable to generalize the metrics proposed in Refs.~\cite{JPmetric,VYS11}.

The existence of three constants of motion makes a metric special, because it allows for the separation of the geodesic equations (the fourth integral of motion is trivially provided by the normalization condition of the 4-momentum of the test particle). Otherwise, geodesic orbits are generally chaotic (see, e.g., Refs.~\cite{BrinkII,chaos}). The Kerr metric possesses such a third constant of motion, the Carter constant, which is associated with a second-rank Killing tensor \cite{Carter68}. Brink~\cite{Brink08,BrinkII,Brink} investigated the existence of a third constant of motion in general stationary, axisymmetric, vacuum spacetimes and showed that for a certain subclass of these metrics a fourth-rank Killing tensor exists. In Newtonian gravity and Maxwell electrostatics, a Carter-like constant exists if the source is axisymmetric and only has even multipole moments that have the same structure as the multipole moments of the Kerr metric~(c.f., Ref.~\cite{WillCarter} and references in Ref.~\cite{WillCarter2}). The general-relativistic analogue of such a matter configuration, however, does not possess a Carter-like constant~\cite{WillCarter2}.

In this paper, I design a new Kerr-like black hole metric which suffers from no pathologies in the exterior domain, admits three independent, exact constants of motion, and depends nonlinearly on four independent deviation functions that measure potential deviations from the Kerr metric in the strong-field regime. This metric is not a solution to the Einstein field equations or any concrete modified theory of gravity, but it can serve as a phenomenological framework for strong-field tests of the no-hair theorem in general classes of gravity theories. 

In order to design a metric with these properties, I carefully introduce arbitrary deviations from the contravariant Kerr metric in a manner that preserves the separability of the Hamilton-Jacobi equations. From this ansatz, I obtain a covariant metric which describes a black hole and which is regular everywhere outside of the event horizon. I further simplify this metric by requiring that it is asymptotically flat, possesses the correct Newtonian limit in the non-relativistic regime, and is consistent with all current weak-field tests in the PPN formalism.

First I design the metric in Boyer-Lindquist-like coordinates and derive a Carter-like constant as well as solutions of the Hamilton-Jacobi equations. By choosing expansions of the deviation functions in suitable power series I write the metric in an explicit form and impose consistency with the current PPN constraints. I then show that the event horizon of this metric is identical to the Kerr event horizon.

I proceed to analyze circular equatorial geodesic orbits in this metric. I derive expressions for the energy, angular momentum, and orbital frequencies of particles on such orbits as a function of the mass, spin, and deviation parameters of the black hole and show that they are significantly modified with respect to their forms in the Kerr metric. From the energy I calculate the location of the ISCO and show that its location is likewise shifted significantly. I further derive expressions of the principal null congruences in this metric and find a transformation of the Boyer-Lindquist-like coordinates to Kerr-Schild-like coordinates. With this transformation I remove the coordinate singularity at the event horizon which allows for a straightforward implementation of fully relativistic magnetohydrodynamic simulations of accretion flows in this metric.

Finally, I relate this metric to known four-dimensional, analytic black hole solutions of modified theories of gravity. I map the new metric in a form appropriately linearized in the deviation parameters or the spin to the black hole solutions in Einstein-Dilaton-Gauss-Bonnet \cite{YS11} and Chern-Simons \cite{YP09} gravity and as well as to the Kerr-like metric of Ref.~\cite{VYS11} written in one of its forms in Ref.~\cite{GY11}. In the latter case, I find exact recursion relations between the deviation parameters of both metrics, which I verify explicitly at the first few nonvanishing orders. I also point out that the black hole solution \cite{braneBH} in Randall-Sundrum-type braneworld gravity \cite{RS2} can be trivially included by choosing the Kerr-Newman metric as the starting point of my analysis.

The paper is organized as follows: In Sec.~\ref{sec:construction}, I design the metric. In Sec.~\ref{sec:properties}, I calculate the locations of the event horizon, Killing horizon, and ergosphere and determine the conditions on the deviations functions so that the exterior domain is regular. I analyze geodesic motion in the equatorial plane and the location of the ISCO in Sec.~\ref{sec:equatorial} and derive the principal null directions and the Kerr-Schild-like form of the metric in Sec.~\ref{sec:KSform}. In Sec.~\ref{sec:mapping}, I map the metric to known metrics from the literature. I formulate my conclusions and discuss astrophysical applications in Sec.~\ref{sec:discussion}. Throughout, I use geometric units, where $G=c=1$, unless I state it explicitly.


\section{Design of a Black Hole Metric with Three Constants of Motion}
\label{sec:construction}

In this section, I design a new class of stationary, axisymmetric, asymptotically flat metrics that describe spinning black holes and that admit three independent constants of motion. The metric elements depend on the mass and spin of the black hole as well as on four free functions that measure potential deviations from the Kerr metric. This class of metrics includes the Kerr metric as the special case if all deviations vanish.

My starting point is the Kerr metric $g_{\rm \mu\nu}^{\rm K}$, which in Boyer-Lindquist coordinates is given by the metric elements
\ba
g_{tt}^{\rm K} &=&-\left(1-\frac{2Mr}{\Sigma}\right), \nonumber \\
g_{t\phi}^{\rm K} &=& -\frac{2Mar\sin^2\theta}{\Sigma}, \nonumber \\
g_{rr}^{\rm K} &=& \frac{\Sigma}{\Delta}, \nonumber \\
g_{\theta \theta}^{\rm K} &=& \Sigma, \nonumber \\
g_{\phi \phi}^{\rm K} &=& \left(r^2+a^2+\frac{2Ma^2r\sin^2\theta}{\Sigma}\right)\sin^2\theta,
\label{kerr}
\ea
where
\ba
\Delta &\equiv& r^2-2Mr+a^2, \nn \\
\Sigma &\equiv& r^2+a^2\cos^2 \theta.
\label{deltasigma}
\ea

In contravariant form, the Kerr metric can be written as (e.g.,~\cite{MTW})
\ba
g_{\rm K}^{\alpha\beta} \frac{\partial}{\partial x^\alpha} \frac{\partial}{\partial x^\beta} = && -\frac{1}{\Delta\Sigma} \left[ (r^2+a^2)\frac{\partial}{\partial t} + a \frac{\partial}{\partial \phi} \right]^2 \nonumber \\
&& + \frac{1}{\Sigma\sin^2\theta} \left[ \frac{\partial}{\partial \phi} + a \sin^2\theta \frac{\partial}{\partial t} \right]^2 \nonumber \\
&& + \frac{\Delta}{\Sigma} \left( \frac{\partial}{\partial r} \right)^2 + \frac{1}{\Sigma} \left( \frac{\partial}{\partial \theta} \right)^2.
\label{eq:contraKerr}
\ea

General stationary and axisymmetric metrics admit two constants of motion, the energy $E$ and axial angular momentum $L_z$, and are of Petrov type I. The Kerr metric, however, is of Petrov type D thanks to the existence of a third constant of motion, the famous Carter constant $Q^{\rm K}$, which Carter~\cite{Carter68} found by explicitly solving the Hamilton-Jacobi equations,
\be
-\frac{\partial S_{\rm K}}{\partial\tau} = \frac{1}{2} g^{\alpha\beta}_{\rm K} \frac{\partial S_{\rm K}}{\partial x^\alpha} \frac{\partial S_{\rm K}}{\partial x^\beta}.
\label{eq:HJKerr}
\ee
As Carter~\cite{Carter68} showed, for a Hamilton-Jacobi function of the form
\be
S_{\rm K} \equiv \frac{1}{2} \mu^2 \tau -E^{\rm K}t +L_z^{\rm K} \phi + S_r^{\rm K}(r) + S_\theta^{\rm K}(\theta),
\ee
where $\tau$ and $\mu$ are the proper time and the rest mass of a test particle on a geodesic orbit in this metric, respectively, the Hamilton-Jacobi equations \eqref{eq:HJKerr} are separable in all four coordinates making geodesic motion in the Kerr metric integrable. I briefly demonstrate the separability of the Hamilton-Jacobi equations below, as these steps will be essential for my design of a new class of Kerr-like metrics.

From Eq.~\eqref{eq:HJKerr}, one obtains the equation
\ba
-\mu^2 = && -\frac{1}{\Delta\Sigma} \left[ -(r^2+a^2)E^{\rm K} + a L_z^{\rm K} \right]^2 \nonumber \\
&& + \frac{1}{\Sigma\sin^2\theta} \left[ L_z^{\rm K} - aE^{\rm K} \sin^2\theta \right]^2 \nonumber \\
&& + \frac{\Delta}{\Sigma} \left( \frac{\partial S_r^{\rm K}}{\partial r} \right)^2 + \frac{1}{\Sigma} \left( \frac{\partial S_\theta^{\rm K}}{\partial \theta} \right)^2.
\ea
After rearranging terms in this equation one can define a separation constant $C^{\rm K}$ such that
\ba
C^{\rm K} = && -r^2\mu^2 \nn \\
&& + \frac{1}{\Delta} \left[ -(r^2+a^2)E^{\rm K} + a L_z^{\rm K} \right]^2 - \Delta \left( \frac{\partial S_r^{\rm K}}{\partial r} \right)^2
\ea
and
\ba
C^{\rm K} = && a^2\mu^2 \cos^2 \theta \nn \\
&& + \frac{1}{\sin^2\theta} \left[ L_z^{\rm K} - aE^{\rm K} \sin^2\theta \right]^2 + \left( \frac{\partial S_\theta^{\rm K}}{\partial \theta} \right)^2.
\ea
From these expressions, it is obvious that the Hamilton-Jacobi equations are separable and one can define the Carter constant as
\be
Q^{\rm K} \equiv C^{\rm K} - (L_z^{\rm K} - aE^{\rm K})^2.
\ee

In the following, I modify the contravariant Kerr metric in such a manner that the corresponding Hamilton-Jacobi equations remain separable. I introduce scalar functions $f(r)$, $g(\theta)$, $A_i(r)$, $i=1,2,5$, and $A_j(\theta)$, $j=3,4,6$, and rewrite Eq.~\eqref{eq:contraKerr} as
\ba
&& g^{\alpha\beta} \frac{\partial}{\partial x^\alpha} \frac{\partial}{\partial x^\beta} \nn \\
&=& -\frac{1}{\Delta\tilde{\Sigma}} \left[ (r^2+a^2)A_1(r)\frac{\partial}{\partial t} + a A_2(r) \frac{\partial}{\partial \phi} \right]^2 \nonumber \\
&& + \frac{1}{\tilde{\Sigma}\sin^2\theta} \left[ A_3(\theta)\frac{\partial}{\partial \phi} + a \sin^2\theta A_4(\theta)\frac{\partial}{\partial t} \right]^2 \nonumber \\
&& + \frac{\Delta}{\tilde{\Sigma}} A_5(r)\left( \frac{\partial}{\partial r} \right)^2 + \frac{1}{\tilde{\Sigma}} A_6(\theta)\left( \frac{\partial}{\partial \theta} \right)^2,
\label{eq:contraKerr_mod}
\ea
where I define
\be
\tilde{\Sigma} \equiv \Sigma + f(r) + g(\theta).
\label{eq:Sigmatilde_gen}
\ee

As in the case of the Kerr metric, I can define a Hamilton-Jacobi function
\be
S \equiv \frac{1}{2} \mu^2 \tau -Et +L_z \phi + S_r(r) + S_\theta(\theta),
\ee
where
\ba
\frac{\partial S}{\partial x^\alpha} &=& p_\alpha, \\
\frac{\partial S}{\partial \tau} &=& \frac{1}{2}\mu^2,
\label{eq:pmu}
\ea
with the corresponding Hamilton-Jacobi equations
\be
-\frac{\partial S}{\partial\tau} = \frac{1}{2} g^{\alpha\beta} \frac{\partial S}{\partial x^\alpha} \frac{\partial S}{\partial x^\beta}.
\label{eq:HJ}
\ee
Here,
\be
p^\alpha \equiv \mu \frac{dx^\alpha}{d\tau}
\label{eq:momentum}
\ee
is the particle's 4-momentum. From Eq.~\eqref{eq:HJ} I obtain the equation
\ba
-\mu^2 = && -\frac{1}{\Delta\tilde{\Sigma}} \left[ -(r^2+a^2)A_1(r)E + a A_2(r)L_z \right]^2 \nonumber \\
&& + \frac{1}{\tilde{\Sigma}\sin^2\theta} \left[ A_3(\theta)L_z - aA_4(\theta)E \sin^2\theta \right]^2 \nonumber \\
&& + \frac{\Delta}{\tilde{\Sigma}} A_5(r)\left( \frac{\partial S_r}{\partial r} \right)^2 + \frac{1}{\tilde{\Sigma}} A_6(\theta)\left( \frac{\partial S_\theta}{\partial \theta} \right)^2.
\ea
Substituting the expression \eqref{eq:Sigmatilde_gen} for $\tilde{\Sigma}$ and rearranging terms I can define a separation constant $C$ such that
\ba
C = && \frac{1}{\Delta} \left[ -(r^2+a^2)A_1(r)E + a A_2(r)L_z \right]^2 \nn \\
&& -\mu^2 \left[ r^2 + f(r) \right] - \Delta A_5(r)\left( \frac{\partial S_r}{\partial r} \right)^2
\label{eq:Kr}
\ea
and
\ba
C = && \frac{1}{\sin^2\theta} \left[ A_3(\theta)L_z - aA_4(\theta)E \sin^2\theta \right]^2 \nn \\
&& + \mu^2 \left[ a^2 \cos^2 \theta + g(\theta) \right] + A_6(\theta)\left( \frac{\partial S_\theta}{\partial \theta} \right)^2.
\label{eq:Ktheta}
\ea

It is now clear that this choice of the functions $f$, $g$, $A_i$, $i=1-6$, preserves the integrability of the Hamilton-Jacobi equations for this metric, because all terms which depend on either the radius $r$ or the polar angle $\theta$ can be separated as in Eqs.~\eqref{eq:Kr} and \eqref{eq:Ktheta}, respectively. Defining the Carter-like constant
\be
Q \equiv C - (L_z - aE)^2,
\ee
I obtain the solutions
\ba
S_r(r) &=& \pm \int dr \frac{1}{\Delta} \sqrt{ \frac{ R(r) }{ A_5(r) } }, \\
S_\theta(\theta) &=& \pm \int d\theta \sqrt{ \frac{ \Theta(\theta) }{ A_6(\theta) } },
\label{eq:S_theta}
\ea
where
\ba
R(r) &\equiv& P^2 \nn \\
&& - \Delta \left\{ \mu^2 \left[ r^2 + f(r) \right] + (L_z - aE)^2 + Q \right\}, 
\label{eq:R(r)}\\
\Theta(\theta) &\equiv& Q + (L_z-aE)^2 -\mu^2 [a^2 \cos^2 \theta + g(\theta)] \nn \\
&& - \frac{1}{\sin^2\theta} \left[ A_3(\theta)L_z - aA_4(\theta)E \sin^2\theta \right]^2, \\
P &\equiv& (r^2 + a^2)A_1(r)E - aA_2(r) L_z.
\label{eq:Theta(theta)}
\ea
From here on I assume that $A_5(r)>0$ and $A_6(\theta)>0$. The former condition is necessary for the regularity of the metric as I will show in the next section, while asymptotic flatness will later require that $A_6(\theta)=1$.

Setting the partial derivatives of the Hamilton-Jacobi function with respect to the constants of motion equal to zero, I derive the following expressions for the proper time and the coordinates $t$ and $\phi$
\ba
\tau &=& \int dr \frac{ r^2 + f(r) }{ \sqrt{A_5(r) R(r)} } + \int d\theta \frac{ a^2 \cos^2 \theta + g(\theta) }{ \sqrt{A_6(\theta) \Theta(\theta)} }, \\
t &=& \int dr \frac{1}{ \Delta \sqrt{A_5(r) R(r)} } \{ a\Delta(L_z-aE) \nn \\
&& + [(r^2+a^2)A_1(r)E-aA_2(r)L_z](r^2+a^2)A_1(r) \} \nn \\
&& + \int d\theta \frac{1}{ \sqrt{A_6(\theta) \Theta(\theta)} } \{ a(aE-L_z) \nn \\
&& +aA_4(\theta)[A_3(\theta)L_z-aA_4(\theta)E\sin^2\theta], \\
\phi &=& \int dr \frac{1}{ \Delta \sqrt{A_5(r) R(r)} } \{ \Delta(L_z-aE) \nn \\
&& + aA_2(r)[(r^2+a^2)A_1(r)E-aA_2(r)L_z] \} \nn \\
&& + \int d\theta \frac{1}{ \sqrt{A_6(\theta) \Theta(\theta)} } \{ aE-L_z \nn \\
&& +A_3(\theta)\csc^2(\theta)[A_3(\theta)L_z-aA_4(\theta)E\sin^2\theta] \}
\ea
as well as the integral relation
\be
\pm \int \frac{ dr }{ \sqrt{A_5(r) R(r)} } = \pm \int \frac{d\theta}{\sqrt{A_6(\theta) \Theta(\theta)}}.
\ee

From the partial derivatives of the Hamilton-Jacobi function with respect to the proper time and the coordinates, I obtain the relations between the momenta $p_\alpha$ and the constants of motion:
\ba
E &=& -p_t,
\label{eq:Econst} \\
L_z &=& p_\phi,
\label{eq:Lzconst} \\
Q &=& A_6(\theta)p_\theta^2 - (L_z-aE)^2 +\mu^2 [a^2 \cos^2 \theta + g(\theta)] \nn \\
&& + \frac{1}{\sin^2\theta} \left[ A_3(\theta)L_z - aA_4(\theta)E \sin^2\theta \right]^2
\label{eq:modCarterConst}
\ea
as well as
\be
p_r = \pm \frac{ \sqrt{R(r)} }{ \Delta \sqrt{A_5(r)} }.
\ee

Finally, using relation~\eqref{eq:momentum}, I find the equations of motion of a particle with rest mass $\mu$:
\ba
\mu \tilde{\Sigma} \frac{dt}{d\tau} &=& -aA_4(\theta) \left[aA_4(\theta)E\sin^2 \theta - A_3(\theta)L_z\right] \nn \\
&& + \frac{(r^2+a^2)A_1(r)}{\Delta} P, 
\label{eq:eom_t}\\
\mu \tilde{\Sigma} \frac{dr}{d\tau} &=& \pm \sqrt{A_5(r) R(r)}, 
\label{eq:eom_r}\\
\mu \tilde{\Sigma} \frac{d\theta}{d\tau} &=& \pm \sqrt{A_6(\theta) \Theta(\theta)}, 
\label{eq:eom_theta}\\
\mu \tilde{\Sigma} \frac{d\phi}{d\tau} &=& -A_3(\theta)\left[ aA_4(\theta)E - A_3(\theta)\frac{L_z}{\sin^2 \theta} \right] \nn \\
&& + \frac{aA_2(r)}{\Delta} P.
\label{eq:eom_phi}
\ea

From the contravariant metric specified in Eq.~\eqref{eq:contraKerr_mod}, I obtain the covariant metric
\ba
g_{tt} &=& -\frac{\tilde{\Sigma}[\Delta A_3(\theta)^2-a^2A_2(r)^2\sin^2\theta]}{[(r^2+a^2)A_1(r)A_3(\theta)-a^2A_2(r)A_4(\theta)\sin^2\theta]^2}, \nn \\
g_{t\phi} &=& -\frac{a[(r^2+a^2)A_1(r)A_2(r)-\Delta A_3(\theta)A_4(\theta)]\tilde{\Sigma}\sin^2\theta}{[(r^2+a^2)A_1(r)A_3(\theta)-a^2A_2(r)A_4(\theta)\sin^2\theta]^2}, \nn \\
g_{rr} &=& \frac{\tilde{\Sigma}}{\Delta A_5(r)}, \nonumber \\
g_{\theta \theta} &=& \frac{\tilde{\Sigma}}{A_6(\theta)}, \nonumber \\
g_{\phi \phi} &=& \frac{\left[(r^2 + a^2)^2 A_1(r)^2 - a^2 \Delta A_4(\theta)^2 \sin^2 \theta \right]}{[(r^2+a^2)A_1(r)A_3(\theta)-a^2A_2(r)A_4(\theta)\sin^2\theta]^2} \nn \\
&& \times \tilde{\Sigma} \sin^2 \theta.
\label{eq:metric_gen}
\ea
This metric reduces smoothly to the Kerr metric in Eq.~\eqref{kerr} if all deviation functions vanish. It is written in Boyer-Lindquist-like coordinates, i.e., in spherical-like coordinates that reduce to Boyer-Lindquist coordinates in the Kerr limit.

The choice of the deviation functions in Eq.~\eqref{eq:contraKerr_mod} may not be the most general one, as one might imagine introducing deviation functions that depend on both the radius and the polar angle such that the separability of the Hamilton-Jacobi equations is still preserved. For massless particles, Eq.~\eqref{eq:contraKerr_mod} could be multiplied by an arbitrary analytic function $A_7(r,\theta)$, which would not spoil the separability of the Hamilton-Jacobi equations since $\mu=0$. However, it is not obvious whether a similar choice can be made for massive particles. Nonetheless, my choice in Eq.~\eqref{eq:contraKerr_mod} is the most general one for deviation functions that depend on either the radius or the polar angle.

Note that the most general stationary, axisymmetric metric in general relativity can be written in terms of only four functions of the radius and polar angle, if a minor technical assumption holds. This assumption is satisfied in particular for stationary, axisymmetric spacetimes that are asymptotically flat and vacuum (see discussion in Ref.~\cite{WaldBook}). Thus, one might expect that some of the deviation functions $f(r)+g(\theta)$, $A_i(r)$ and $A_j(\theta)$ are either trivially related or equal to unity. Indeed, as I will show below, requiring that the metric is asymptotically flat reduces the number of independent deviation functions to four.

In order to obtain an explicit form of this metric, I write the deviation functions $A_i(r)$, $i=1,2,5$, as a power series in $M/r$,
\be
A_i(r) \equiv \sum_{n=0}^\infty \alpha_{in} \left( \frac{M}{r} \right)^n,~~~~i=1,2,5,
\ee
as well as
\ba
f(r) &\equiv& \sum_{n=0}\epsilon_n \frac{M^n}{r^{n-2}}, \\
g(\theta) &\equiv& M^2 \sum_{k,l=0}^\infty \gamma_{kl} \sin^k\theta \cos^l\theta,
\ea
where in the last expression $\gamma_{00}=0$.

In Boyer-Lindquist-like coordinates, asymptotically flat metrics must be of the form (e.g., \cite{Heusler96}, but see discussion in Ref.~\cite{WaldBook})
\ba
ds^2 &=& -\left[ 1-\frac{2M}{r} + \mathcal{O}\left(r^{-2}\right) \right]dt^2 
\nonumber \\
&&- \left[\frac{4Ma}{r}\sin^2\theta + \mathcal{O}\left(r^{-2}\right) \right]dtd\phi + \bigg[1 + \mathcal{O}\left(r^{-1}\right) \bigg]
\nonumber \\
&&\times \bigg[dr^2 + r^2\left(d\theta^2 + \sin^2\theta d\phi^2\right) \bigg].
\label{eq:asympt}
\ea

Expanding the metric given by Eq.~\eqref{eq:metric_gen} in $1/r$, it follows that $\alpha_{10}=\alpha_{20}=\alpha_{50}=1$ as well as $A_3(\theta)=A_4(\theta)=A_6(\theta)=1$. Further, I set $\alpha_{11}=\alpha_{21}=\alpha_{51}=0$, which defines the parameter $M$ as the mass of the central object. Otherwise, the mass would have to be rescaled appropriately. This choice also defines the parameter $a$ as the spin of the central object.

The deviation parameters can be further constrained in the PPN framework \cite{WillLRR}. The general PPN metric can be written in the form
\be
ds^2 = -A_{\rm PPN}(r)dt^2 + B_{\rm PPN}(r)dr^2 + r^2d\Omega,
\ee
where
\ba
A_{\rm PPN}(r) &\equiv& 1-\frac{2M}{r}+2(\beta_{\rm PPN}-\gamma_{\rm PPN})\frac{M^2}{r^2}, 
\label{eq:PPN_A}\\
B_{\rm PPN}(r) &\equiv& 1+2\gamma_{\rm PPN}\frac{M}{r}, 
\label{eq:PPN_B}\\
d\Omega &\equiv& d\theta^2 + \sin^2\theta d\phi^2.
\ea
In general relativity, $\beta_{\rm PPN}=\gamma_{\rm PPN}=1$. Comparing this metric with the expansion of the metric given by Eq.~\eqref{eq:metric_gen} in $1/r$, I obtain the relations
\ba
2(\beta_{\rm PPN} - \gamma_{\rm PPN}) &=& \epsilon_2 -2\alpha_{12} + \frac{g(\theta)}{M^2}, 
\label{eq:PPNnum1}\\
\gamma_{\rm PPN} &=& 1,
\ea
which implies the equation
\be
\beta_{\rm PPN} -1 = \frac{1}{2}\left[ \epsilon_2 -2\alpha_{12} + \frac{g(\theta)}{M^2} \right].
\label{eq:betaconstraint}
\ee
This quantity is tightly constrained by observations \cite{PPNconstraint}:
\be
\left|\beta_{\rm PPN} -1 \right| \leq 2.3\times10^{-4}.
\ee
In order to avoid any fine-tuning between the parameters $\epsilon_2$ and $\alpha_{12}$ and the function $g(\theta)$, I will also set $\epsilon_2=\alpha_{12}=g(\theta)=0$. Other choices for these parameters that satisfy Eq.~\eqref{eq:betaconstraint} could also be made, such as setting $\epsilon_2=2\alpha_{12}$ and $g(\theta)=0$, but I will exclude them here for simplicity.

Summarizing these results, the newly designed metric is given by the elements
\ba
g_{tt} &=& -\frac{\tilde{\Sigma}[\Delta-a^2A_2(r)^2\sin^2\theta]}{[(r^2+a^2)A_1(r)-a^2A_2(r)\sin^2\theta]^2}, \nn \\
g_{t\phi} &=& -\frac{a[(r^2+a^2)A_1(r)A_2(r)-\Delta]\tilde{\Sigma}\sin^2\theta}{[(r^2+a^2)A_1(r)-a^2A_2(r)\sin^2\theta]^2}, \nn \\
g_{rr} &=& \frac{\tilde{\Sigma}}{\Delta A_5(r)}, \nn \\
g_{\theta \theta} &=& \tilde{\Sigma}, \nn \\
g_{\phi \phi} &=& \frac{\tilde{\Sigma} \sin^2 \theta \left[(r^2 + a^2)^2 A_1(r)^2 - a^2 \Delta \sin^2 \theta \right]}{[(r^2+a^2)A_1(r)-a^2A_2(r)\sin^2\theta]^2},
\label{eq:metric}
\ea
where
\ba
A_1(r) &=& 1 + \sum_{n=3}^\infty \alpha_{1n} \left( \frac{M}{r} \right)^n, 
\label{eq:A1}\\
A_2(r) &=& 1 + \sum_{n=2}^\infty \alpha_{2n} \left( \frac{M}{r} \right)^n, 
\label{eq:A2}\\
A_5(r) &=& 1 + \sum_{n=2}^\infty \alpha_{5n} \left( \frac{M}{r} \right)^n, 
\label{eq:A5}\\
\tilde{\Sigma} &=& r^2 + a^2 \cos^2\theta + f(r), 
\label{eq:Sigmatilde}\\
f(r) &=& \sum_{n=3}^\infty\epsilon_n \frac{M^n}{r^{n-2}}.
\label{eq:f}
\ea
This metric is asymptotically flat, has the correct Newtonian limit, and is consistent with the current PPN constraints. At lowest order, i.e., truncating the series in Eqs.~\eqref{eq:A1}--\eqref{eq:f} at the first nonvanishing order in the deviation parameters, this metric depends on four parameters in addition to the mass $M$ and the spin $a$: $\alpha_{13}$, $\alpha_{22}$, $\alpha_{52}$, and $\epsilon_3$.


\section{Metric Properties}
\label{sec:properties}

In this section, I analyze some of the properties of the newly designed metric and calculate the location of the event horizon. I likewise determine conditions which the deviation functions in Eqs.~\eqref{eq:A1}--\eqref{eq:f} must satisfy so that the exterior domain is regular.

\subsection{Event horizon, Killing horizon, and ergosphere}

First I find the event horizon. My analysis is similar to the one in Ref.~\cite{Joh13}. For a stationary, axisymmetric, asymptotically flat metric of the kind given by Eq.~\eqref{eq:metric}, the event horizon, if one exists, is located at a radius
\be
r_{\rm hor} \equiv H(\theta).
\ee
The function $H(\theta)$ is a solution of the ordinary differential equation \cite{Joh13,Thornburg07}
\be
g^{rr} + g^{\theta\theta} \left( \frac{dH}{d\theta} \right)^2 = 0,
\label{eq:hor_master}
\ee
where, from Eq.~\eqref{eq:metric}, $g^{rr}=1/g_{rr}$ and $g^{\theta\theta}=1/g_{\theta\theta}$.

As shown in Ref.~\cite{Joh13}, at the poles and in the equatorial plane this equation reduces to
\be
g^{rr} = 0
\label{eq:hor_axieq}
\ee
due to axi- and reflection symmetry. From the $g_{rr}$ and $g_{\theta\theta}$ elements of the metric in Eq.~\eqref{eq:metric}, it is obvious that
\ba
\tilde{\Sigma}&>&0, 
\label{eq:Sigmatilde_cond} \\
A_5(r)&>&0
\label{eq:A5_cond}
\ea
at all radii $r\geq r_{\rm hor}$, because otherwise the metric signature would change at some radius outside of the event horizon which would introduce a singularity at that location (see the discussion in the next subsection). Therefore, Eq.~\eqref{eq:hor_axieq} reduces to the condition
\be
\Delta = 0
\ee
both at the poles and in the equatorial plane, which means that at these locations the horizon coincides with the horizon $r_+$ of a Kerr black hole,
\be
r_+ \equiv H_K \equiv M + \sqrt{M^2-a^2}.
\label{eq:Kerrhor}
\ee
As in Ref.~\cite{Joh13}, here I use both notations $r_+$ and $H_K$ for the Kerr horizon. While the former is typically used, the latter is often used in the numerical relativity literature.

For arbitrary angles $\theta$, the horizon equation \eqref{eq:hor_master} takes the form
\be
\frac{1}{\tilde{\Sigma}} \left[ \Delta A_5(H) + \left( \frac{dH}{d\theta} \right)^2 \right] = 0,
\ee
which reduces to the equation
\be
\Delta A_5(H) + \left( \frac{dH}{d\theta} \right)^2 = 0
\label{eq:genmetrichor}
\ee
because $\tilde{\Sigma}>0$. Since also $A_5(H)>0$, it is obvious that the Kerr horizon given by Eq.~\eqref{eq:Kerrhor} is always a solution of Eq.~\eqref{eq:genmetrichor}. Furthermore, should a different solution to the horizon equation exist, it must lie inside of the radius $r_+$ at angles $0<\theta<\pi/2$, $\pi/2<\theta<\pi$, because the second-term in Eq.~\eqref{eq:genmetrichor} is non-negative and $A_5>0$, which implies that a solution for which $\Delta\neq0$ must have $\Delta<0$ and, therefore, $r<r_+$. From this analysis I conclude that the event horizon of the metric in Eq.~\eqref{eq:metric} is the Kerr event horizon.

The Killing horizon of the metric given by Eq.~\eqref{eq:metric} is located at the roots of the equation
\be
g_{t\phi}^2 - g_{tt}g_{\phi\phi} = \frac{ \Delta \tilde{\Sigma}^2 \sin^2\theta }{ \left[ \left(r^2+a^2\right)A_1(r) - a^2 A_2(r) \sin^2\theta \right]^2 },
\label{eq:Killinghor1}
\ee
which is equivalent to the equation
\be
\Delta = 0
\ee
as long as the denominator of Eq.~\eqref{eq:Killinghor1} is nonzero at and outside of the Killing horizon. I will return to this requirement below. Consequently, the Killing horizon coincides with the event horizon. This suggests that a generalized version of Hawking's rigidity theorem \cite{rigidity} may hold for this metric.

The ergosphere is determined by solving the equation
\be
g_{tt} = 0,
\ee
which reduces to the equation
\be
\Delta - a^2 A_2(r)^2 \sin^2\theta = 0.
\label{eq:ergosphere}
\ee
The shape of the ergosphere, therefore, depends on the spin and on the function $A_2(r)$. 

\begin{figure}[ht]
\begin{center}
\psfig{figure=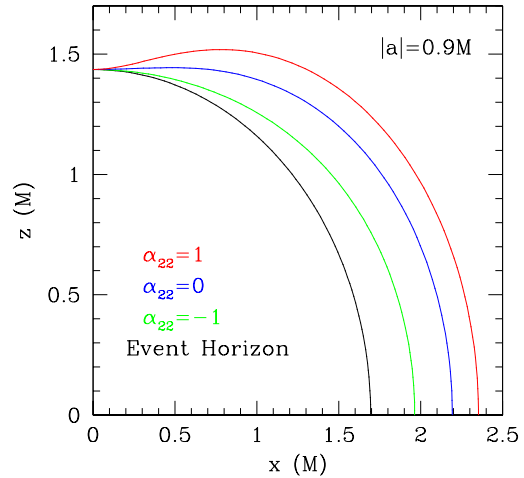,height=2.93in}
\end{center}
\caption{Event horizon and ergosphere in the $xz$-plane of a black hole with spin $|a|=0.9M$ for several values of the parameter $\alpha_{22}$. The event horizon coincides with the horizon of a Kerr black hole and depends only on the spin $a$. For increasing values of the parameter $\alpha_{22}$, the extent of the ergosphere increases at angles $0<\theta<\pi$. At the poles, the event horizon and ergosphere coincide irrespectively of the value of the parameter $\alpha_{22}$.}
\label{fig:ergo}
\end{figure}

At lowest nonvanishing order, Eq.~\eqref{eq:ergosphere} reduces to the equation
\be
r^4 \Delta - a^2 (r^2 + \alpha_{22}M^2)^2 \sin^2\theta = 0,
\ee
which has to be solved numerically. In Fig.~\ref{fig:ergo}, I plot the event horizon and the ergosphere of a black hole with a spin $|a|=0.9M$ for several values of the parameter $\alpha_{22}$ in the $xz$-plane, where $x\equiv \sqrt{r^2+a^2}\sin\theta$, $z\equiv r\cos\theta$. For increasing values of the parameter $\alpha_{22}$, the extent of the ergosphere increases.

\subsection{Regularity of the exterior domain}

Following the analysis in Ref.~\cite{Joh13}, I analyze the exterior domain of the metric given by Eq.~\eqref{eq:metric} for the existence of any singularities or pathological regions, which might have been introduced by the deviation functions in Eqs.~\eqref{eq:A1}--\eqref{eq:f}. Arbitrary deviations from the Kerr metric can lead to a violation of Lorentzian signature or the existence of closed timelike curves. In the former case, the determinant of the metric is no longer negative definite and, in the latter case, the $(\phi,\phi)$ element of the metric is negative.

The determinant of the metric in Eq.~\eqref{eq:metric} is given by the expression
\be
\det\left( g_{\alpha\beta} \right) = - \frac{\tilde{\Sigma}^4 \sin^2\theta}{A_5(r) \left[ A_1(r) (r^2+a^2) - a^2 A_2(r) \sin^2\theta \right]^2}.
\label{eq:gendet}
\ee
For the determinant to be negative definite, I impose the requirement
\be
A_1(r) (r^2+a^2) - a^2 A_2(r) \sin^2\theta \neq 0,
\ee
in addition to the conditions in Eqs.~\eqref{eq:Sigmatilde_cond} and \eqref{eq:A5_cond}, which must be fulfilled everywhere on and outside of the event horizon.

At the lowest order in the deviation parameters, I can write these conditions in the following manner:
\ba
\alpha_{13} &\neq& \frac{ a^2 r (r^2+\alpha_{22}M^2)\sin^2\theta - r^3(r^2+a^2) }{ M^3 (r^2+a^2) },
\label{eq:alpha13alpha22} \\
\alpha_{52} &>& - \frac{ \left( M+\sqrt{M^2-a^2} \right)^2 }{M^2},
\label{eq:alpha52cond} \\
\epsilon_3 &>& - \frac{ \left( M+\sqrt{M^2-a^2} \right)^3 }{M^3}.
\label{eq:ep3cond}
\ea
In these expressions I replaced the radius with the Kerr horizon radius $r_+$ and, thereby, obtain a lower bound on the deviation parameters, which is valid at all radii $r\geq r_+$.

\begin{figure}[ht]
\begin{center}
\psfig{figure=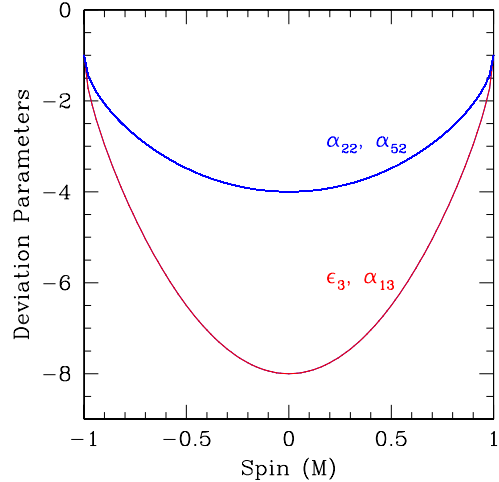,height=2.93in}
\end{center}
\caption{Lower bounds on the deviation parameters $\alpha_{13}$, $\alpha_{22}$, $\alpha_{52}$, and $\epsilon_3$ of the metric in Eq.~\eqref{eq:metric} at the lowest nonvanishing order of the deviation parameters. For values of the deviation parameters greater than these bounds, the event horizon and the exterior domain are regular, i.e., free of curvature singularities and closed timelike curves.}
\label{fig:parameterspace}
\end{figure}

In order to exclude the existence of closed timelike curves in the exterior domain, I derive from the $(\phi,\phi)$ element of the metric the additional requirement
\be
A_1(r)^2\left(r^2+a^2\right)^2 - a^2 \Delta \sin^2\theta > 0
\label{eq:A1cond}
\ee
or, at lowest order,
\be
\alpha_{13} > - \frac{ \left( M+\sqrt{M^2-a^2} \right)^3 }{ M^3 }
\label{eq:alpha13cond}
\ee
on and outside of the horizon, where in the last equation I, again, inserted the Kerr horizon radius $r=r_+$. Combining Eq.~\eqref{eq:alpha13cond} with the condition given by Eq.~\eqref{eq:alpha13alpha22}, I obtain the requirement
\be
\alpha_{22} > - \frac{ \left( M+\sqrt{M^2-a^2} \right)^2 }{ M^2 }.
\label{eq:alpha22cond}
\ee
Therefore, Eqs.~\eqref{eq:alpha52cond}, \eqref{eq:ep3cond}, \eqref{eq:alpha13cond}, and \eqref{eq:alpha22cond} define the allowed ranges of the deviation parameters in the lowest-order metric as a function of the spin. I plot these functions in Fig.~\ref{fig:parameterspace}. At higher values of the spin, the lower limit on the deviation functions is more tight than at smaller spin values. Note, however, that at maximal spin $a=\pm M$ the lower limit on all four parameters is $-1$, which means that even in this case both positive and negative deviations from the Kerr metric can be studied. This is different in, e.g., the metric designed in Ref.~\cite{JPmetric}, where the metric properties can change drastically for positive deviations larger than a lower bound that approaches zero in the limit $a\rightarrow \pm M$ \cite{JPmetric,Joh13}.

I verified numerically the absence of curvature singularities at the horizon located at $r=r_+$ for a large number of different values of the spin in the range $-1\leq a/M \leq1$ and of the deviation parameters $\epsilon_3,~\alpha_{13},~\alpha_{22},~\alpha_{52}$ in the allowed parameter space by evaluating the Kretschmann scalar,
\be
K \equiv R_{\alpha\beta\gamma\delta} R^{\alpha\beta\gamma\delta}
\label{Kretschmann}
\ee
at polar angles $\theta=\pi/2-5\pi n/21$, $n=0,1,2$, where $R^{\alpha}_{~\beta\gamma\delta}$ is the Riemann tensor. For this analysis, I chose equidistant values of the spin in steps of $\Delta a=0.1M$ and values of the deviations parameters ranging from $-3$ to $5$ in steps of $1$ neglecting any parameter combinations that lie in the excluded part of the parameter space. In all cases, the event horizon is the location of a coordinate singularity as in the Kerr metric. In Sec.~\ref{sec:KSform}, I will construct an explicit coordinate transformation that removes the coordinate singularity at the event horizon, proving that it is indeed the location of a coordinate singularity. I, therefore, conclude that the metric in Eq.~\eqref{eq:metric} harbors a black hole with a regular exterior domain.

\section{Keplerian Frequency, Energy, Angular Momentum}
\label{sec:equatorial}

\begin{figure*}[ht]
\begin{center}
\psfig{figure=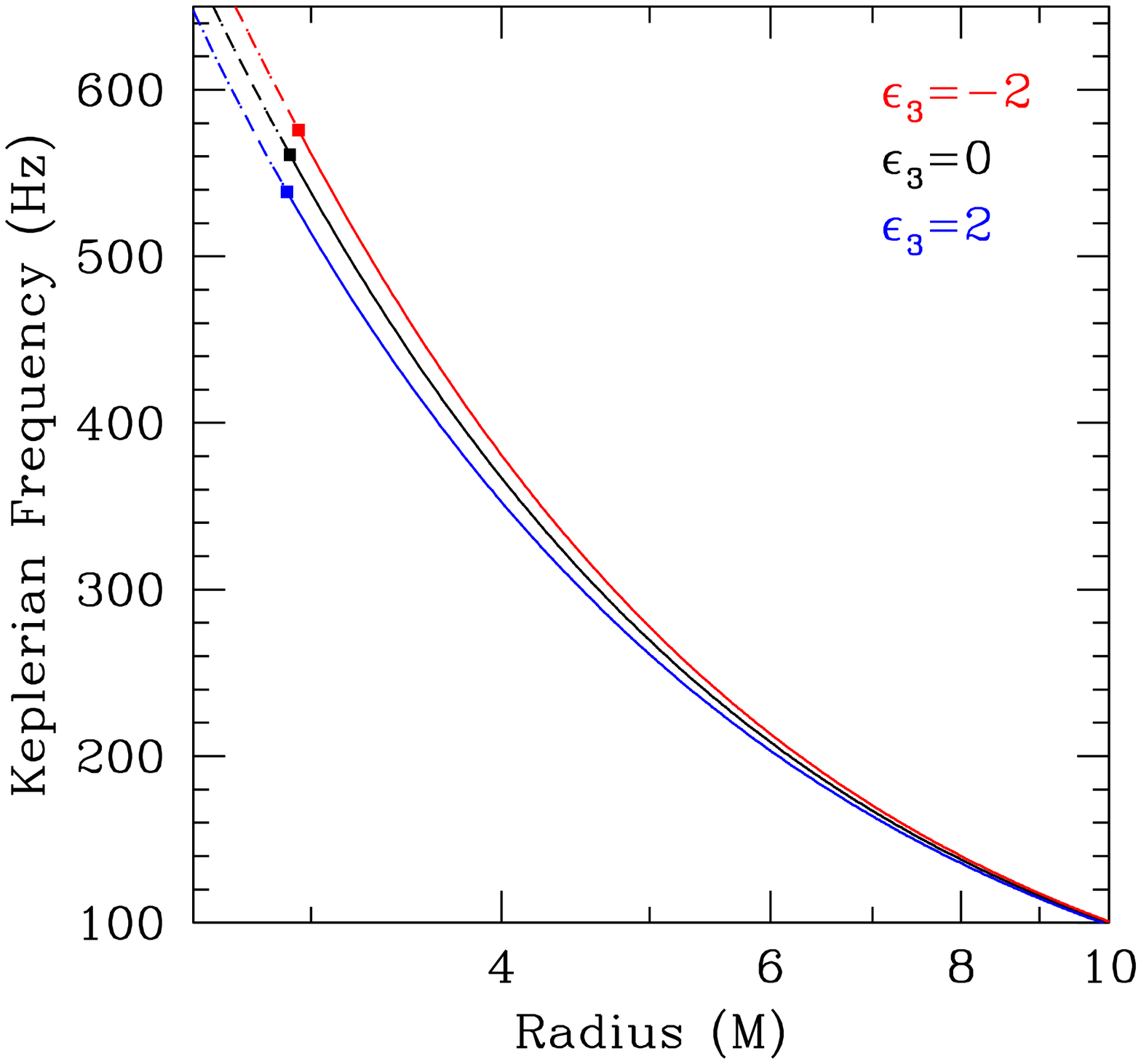,width=0.32\textwidth}
\psfig{figure=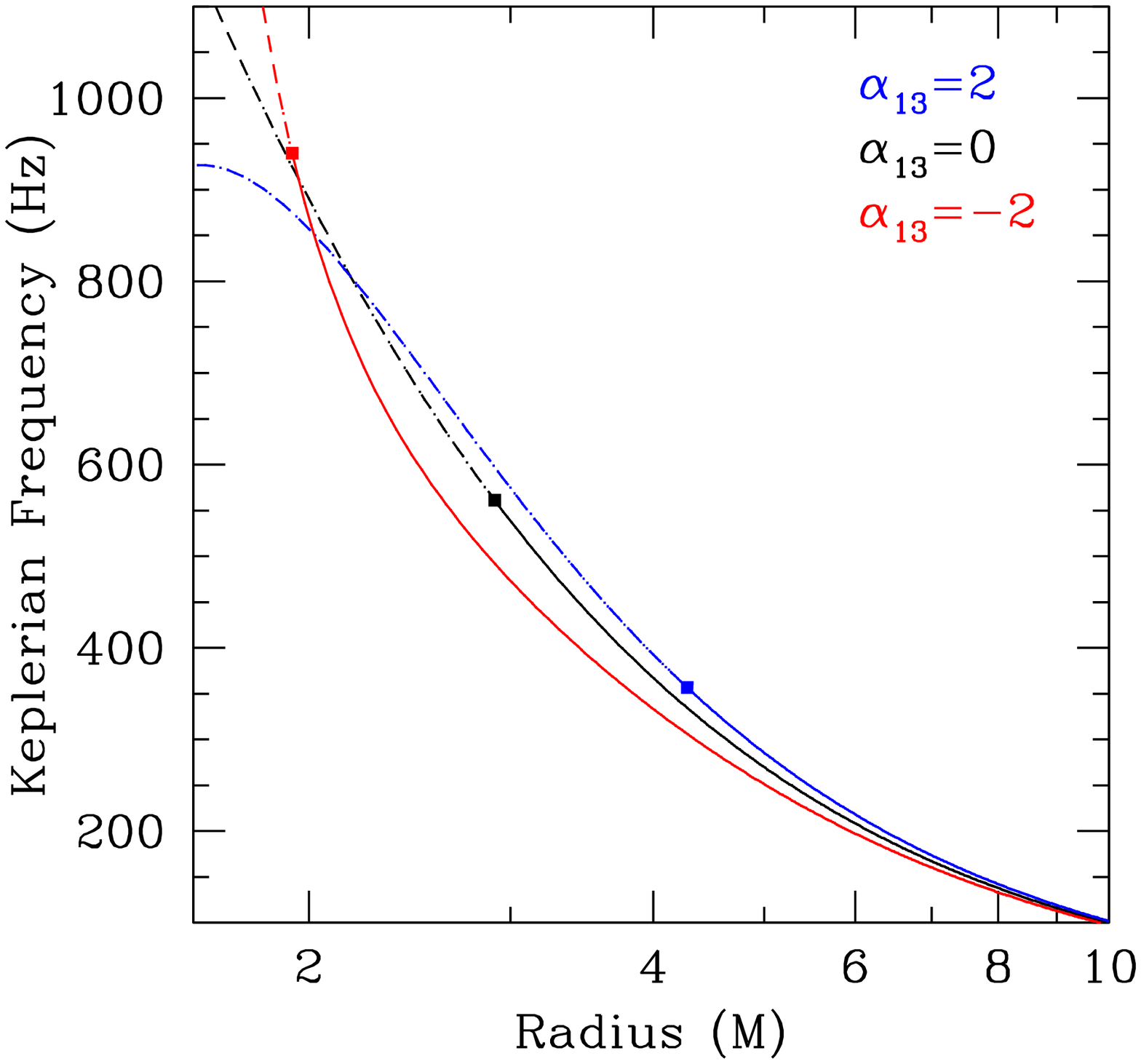,width=0.32\textwidth}
\psfig{figure=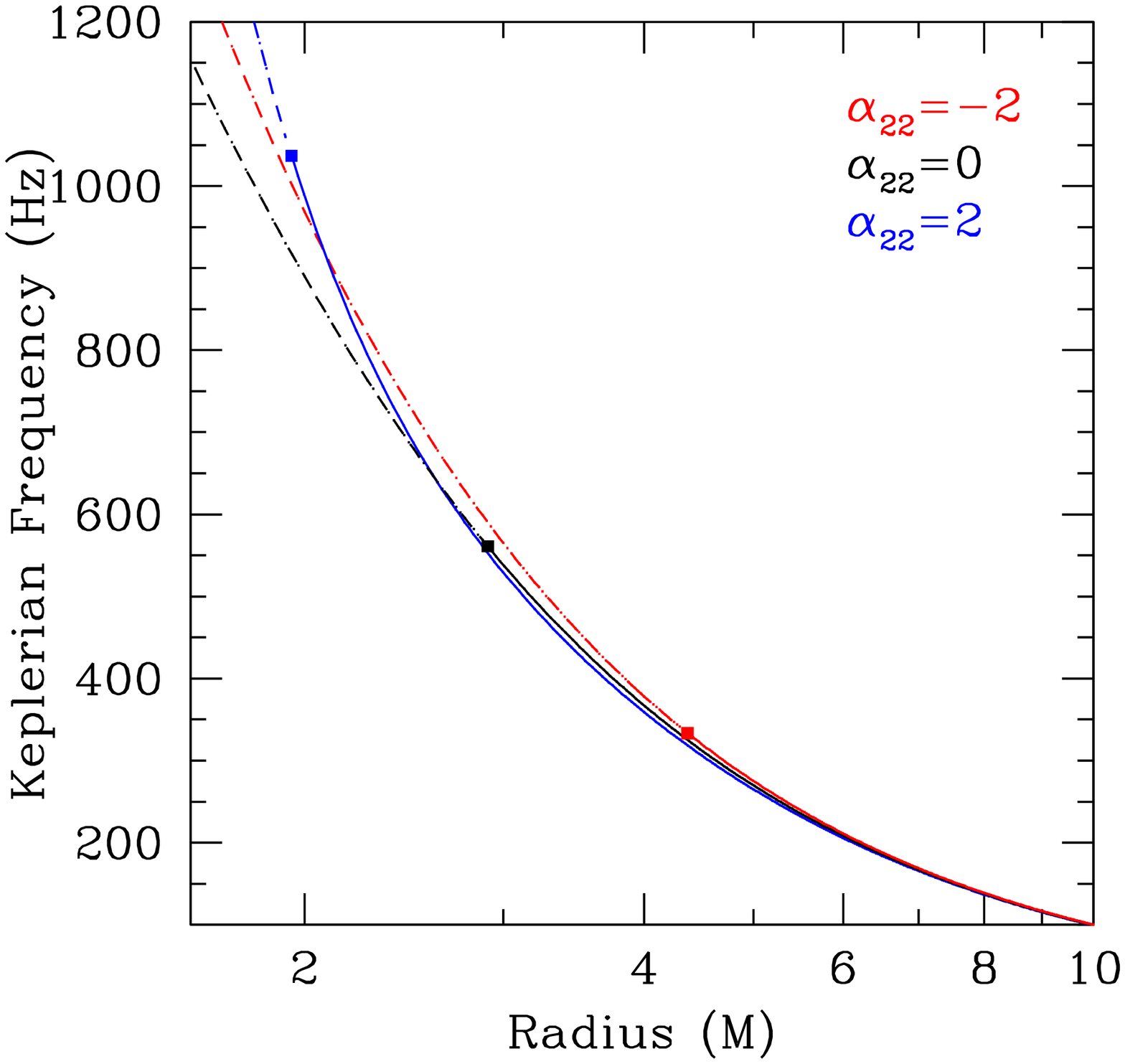,width=0.32\textwidth}
\end{center}
\caption{Dependence of the Keplerian frequency $\nu_\phi= c^3 \Omega_\phi/2\pi GM$ on the deviation parameters $\epsilon_3$, $\alpha_{13}$, and $\alpha_{22}$. Each panel shows the Keplerian frequency $\nu_\phi$ as a function of radius for a black hole with mass $M=10M_\odot$ and spin $a=0.8M$ for different values of one deviation parameter while setting the other two equal to zero. At a given radius, the Keplerian frequency increases for decreasing values of the parameters $\epsilon_3$ and $\alpha_{22}$ and for increasing values of the parameter $\alpha_{13}$. The dot denotes the location of the ISCO.}
\label{fig:kep}
\end{figure*}

In this section, I derive expressions for the dynamical frequencies, energy, and axial angular momentum of a particle on a circular equatorial orbit. My derivation follows closely the one in Refs.~\cite{Ryan95,Gair08}. Then I proceed to calculate the radius of the ISCO.

The geodesic equations,
\be
\frac{d^2 x^\alpha}{d\tau^2} = -\Gamma^\alpha_{\beta\gamma} \frac{dx^\beta}{d\tau}\frac{dx^\gamma}{d\tau},
\ee
where $\Gamma^\alpha_{\beta\gamma}$ are the Christoffel symbols, can also be written in the form
\be
\frac{d}{d\tau} \left( g_{\delta\alpha} \frac{dx^\alpha}{d\tau} \right) = \frac{1}{2} \frac{\partial g_{\beta\gamma} }{\partial x^\delta} \frac{dx^\beta}{d\tau}\frac{dx^\gamma}{d\tau}.
\ee
Due to axi- and reflection symmetry, $dr/d\tau=d\theta/d\tau=d^2r/d\tau^2=0$ for particles on circular equatorial orbits. Therefore, the equation that governs particle motion in the radial direction reduces to the relation
\be
\partial_r g_{tt}\dot{t}^2 + 2\partial_r g_{t\phi} \dot{t}\dot{\phi} + \partial_r g_{\phi\phi} \dot{\phi}^2 = 0,
\label{eq:kepeq}
\ee
where $\partial_r \equiv \partial/\partial r$ and the dot denotes the derivative $d/d\tau$. From this equation, the Keplerian frequency $\Omega_\phi \equiv \dot{\phi}/\dot{t}=p_\phi/p_t$ as observed at radial infinity can be expressed in the form
\be
\Omega_\phi = \frac{ -\partial_r g_{t\phi} \pm \sqrt{ (\partial_r g_{t\phi})^2 - \partial_r g_{tt} \partial_r g_{\phi\phi} } }{ \partial_r g_{\phi\phi} },
\label{eq:keplerian}
\ee
where the upper sign refers to prograde orbits, while the lower sign refers to retrograde orbits. Comparing this expression with the metric in Eq.~\eqref{eq:metric}, it is obvious that the Keplerian frequency is independent of the function $A_5(r)$, because it is independent of the metric element $g_{rr}$.

Eq.~\eqref{eq:keplerian} assumes that $\partial_r g_{\phi\phi}\neq0$. At least in the case of the lowest-order metric, for large negative values of the deviation parameter $\alpha_{13}$ near the lower bound given by Eq.~\eqref{eq:alpha13cond}, the term $\partial_r g_{\phi\phi}$ can be either zero or negative near but outside of the event horizon; otherwise it is always positive. If the term $\partial_r g_{\phi\phi}$ vanishes, the Keplerian frequency is simply $\Omega_\phi=-\partial_r g_{tt}/2\partial_r g_{t\phi}$ from Eq.~\eqref{eq:kepeq}.

In Fig.~\ref{fig:kep}, I illustrate the dependence of the Keplerian frequency on the three lowest-order deviation parameters $\epsilon_3$, $\alpha_{13}$, and $\alpha_{22}$. I plot the Keplerian frequency for a black hole with mass $M=10M_\odot$ and spin $a=0.8M$ as a function of the radius for different values of one of these parameters while setting the other two equal to zero. At a given radius, the Keplerian frequency increases for decreasing values of the parameters $\epsilon_3$ and $\alpha_{22}$, while the Keplerian frequency increases for increasing values of the parameter $\alpha_{13}$.

From the equation
\be
p^\alpha p_\alpha = -\mu^2
\ee
evaluated in the equatorial plane, where $p^\alpha$ is given by Eq.~\eqref{eq:momentum}, I obtain the effective potential
\ba
&& V_{\rm eff}(r) \equiv \frac{1}{2}\mu^2\left( g_{rr}\dot{r}^2 + g_{\theta\theta} \dot{\theta}^2 \right)\nn \\
&& = - \frac{1}{2} (g^{tt}E^2 -2g^{t\phi}EL_z + g^{\phi\phi}L_z^2 + \mu^2).
\ea
In this expression, I substituted the constants of motion $E$ and $L_z$ given by Eqs.~\eqref{eq:Econst}--\eqref{eq:Lzconst} for the momentum components $p_t$ and $p_\phi$. Circular equatorial orbits are governed by the equations
\ba
V_{\rm eff}(r) &=& 0, 
\label{eq:Veff=0} \\
\frac{d V_{\rm eff}(r)}{dr} &=& 0.
\label{eq:dVeff=0}
\ea

In terms of the constants of motion, I write the Keplerian frequency as
\be
\Omega_\phi = \frac{p_\phi}{p_t} = - \frac{ g_{t\phi}E + g_{tt}L_z }{ g_{\phi\phi}E + g_{t\phi}L_z }.
\ee
Combining this equation with Eqs.~\eqref{eq:keplerian} and \eqref{eq:Veff=0}, I can solve for the energy and axial angular momentum and obtain the expressions
\ba
\frac{E}{\mu} &=& - \frac{ g_{tt} + g_{t\phi}\Omega_\phi }{ \sqrt{ -g_{tt}-2g_{t\phi}\Omega_\phi-g_{\phi\phi}\Omega_\phi^2 } }, 
\label{eq:E} \\
\frac{L_z}{\mu} &=& \pm \frac{ g_{t\phi} + g_{\phi\phi}\Omega_\phi }{ \sqrt{ -g_{tt}-2g_{t\phi}\Omega_\phi-g_{\phi\phi}\Omega_\phi^2 } },
\label{eq:Lz}
\ea
where, again, the upper and lower signs refer to prograde and retrograde orbits, respectively. These expressions are likewise independent of the deviation function $A_5(r)$.

\begin{figure*}[ht]
\begin{center}
\psfig{figure=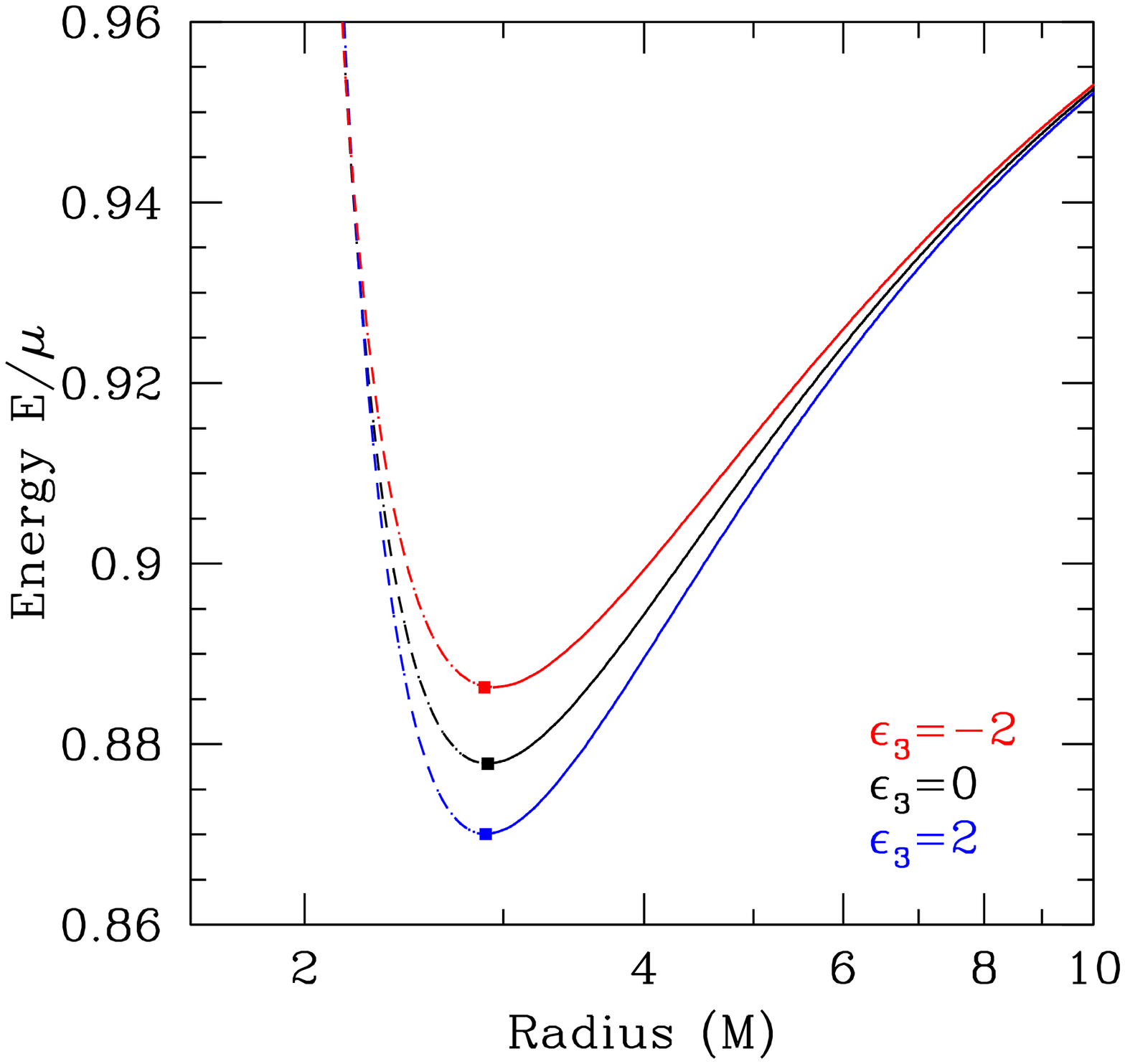,width=0.32\textwidth}
\psfig{figure=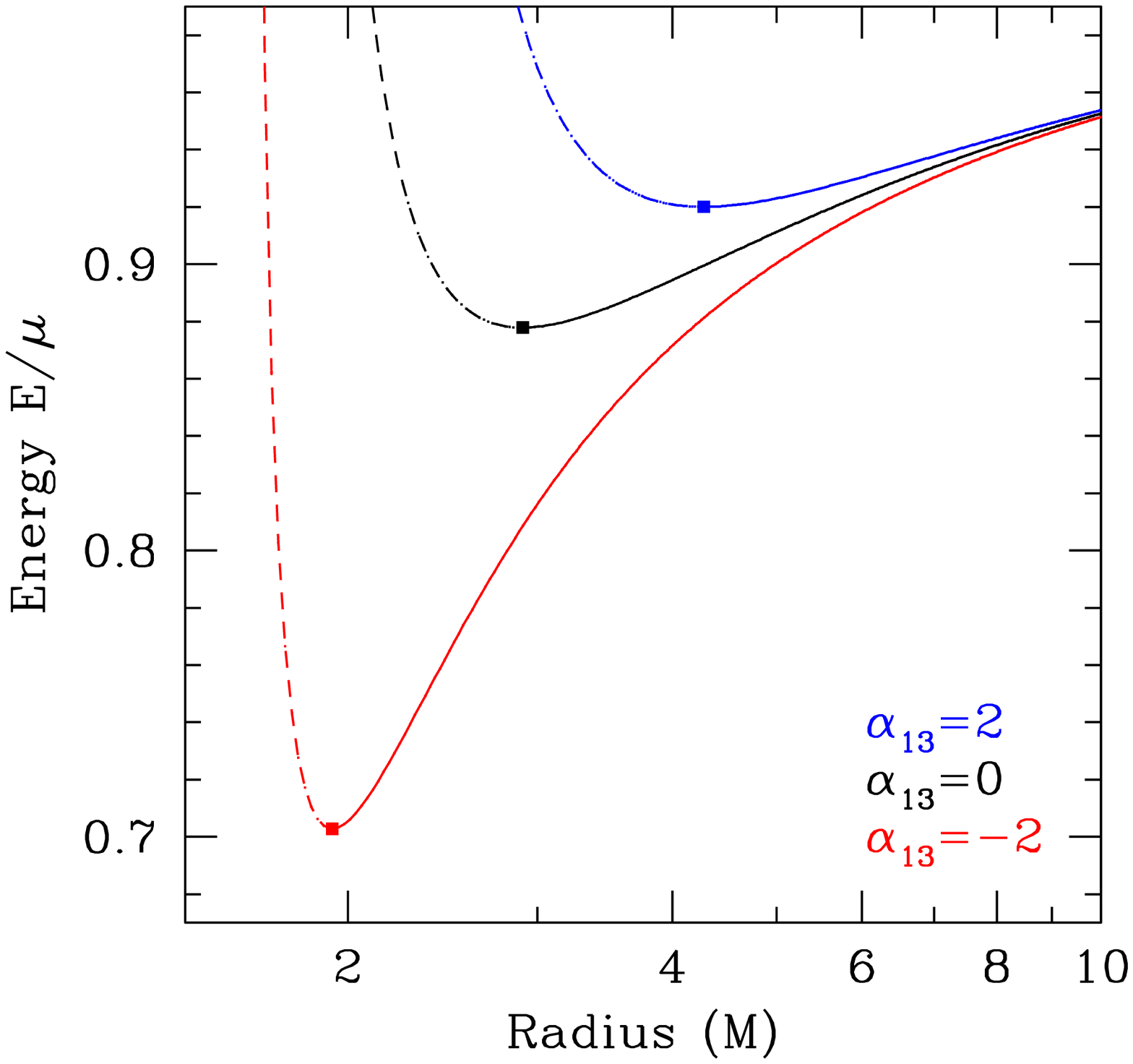,width=0.32\textwidth}
\psfig{figure=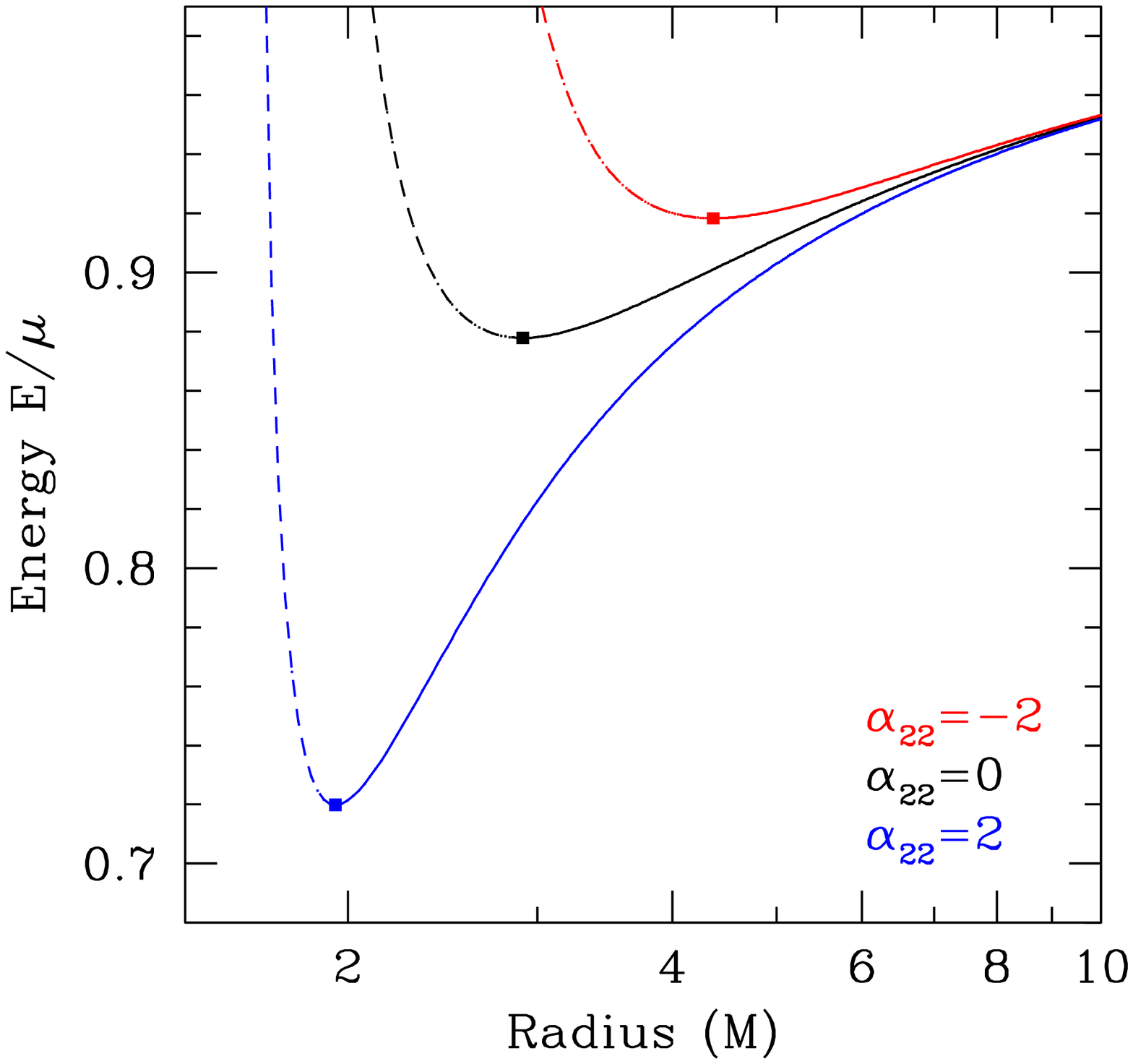,width=0.32\textwidth}
\end{center}
\caption{Dependence of the energy $E$ of a particle with rest mass $\mu$ on a circular equatorial orbit around a black hole with spin $a=0.8M$ as observed at infinity on the deviation parameters $\epsilon_3$, $\alpha_{13}$, and $\alpha_{22}$. Each panel shows the energy as a function of radius for different values of one deviation parameter while setting the other two equal to zero. At a given radius, the energy increases for decreasing values of the parameters $\epsilon_3$ and $\alpha_{22}$ and for increasing values of the parameter $\alpha_{13}$. The ISCO is located at the minimum of the energy and is denoted by a dot.}
\label{fig:en}
\end{figure*}

\begin{figure*}[ht]
\begin{center}
\psfig{figure=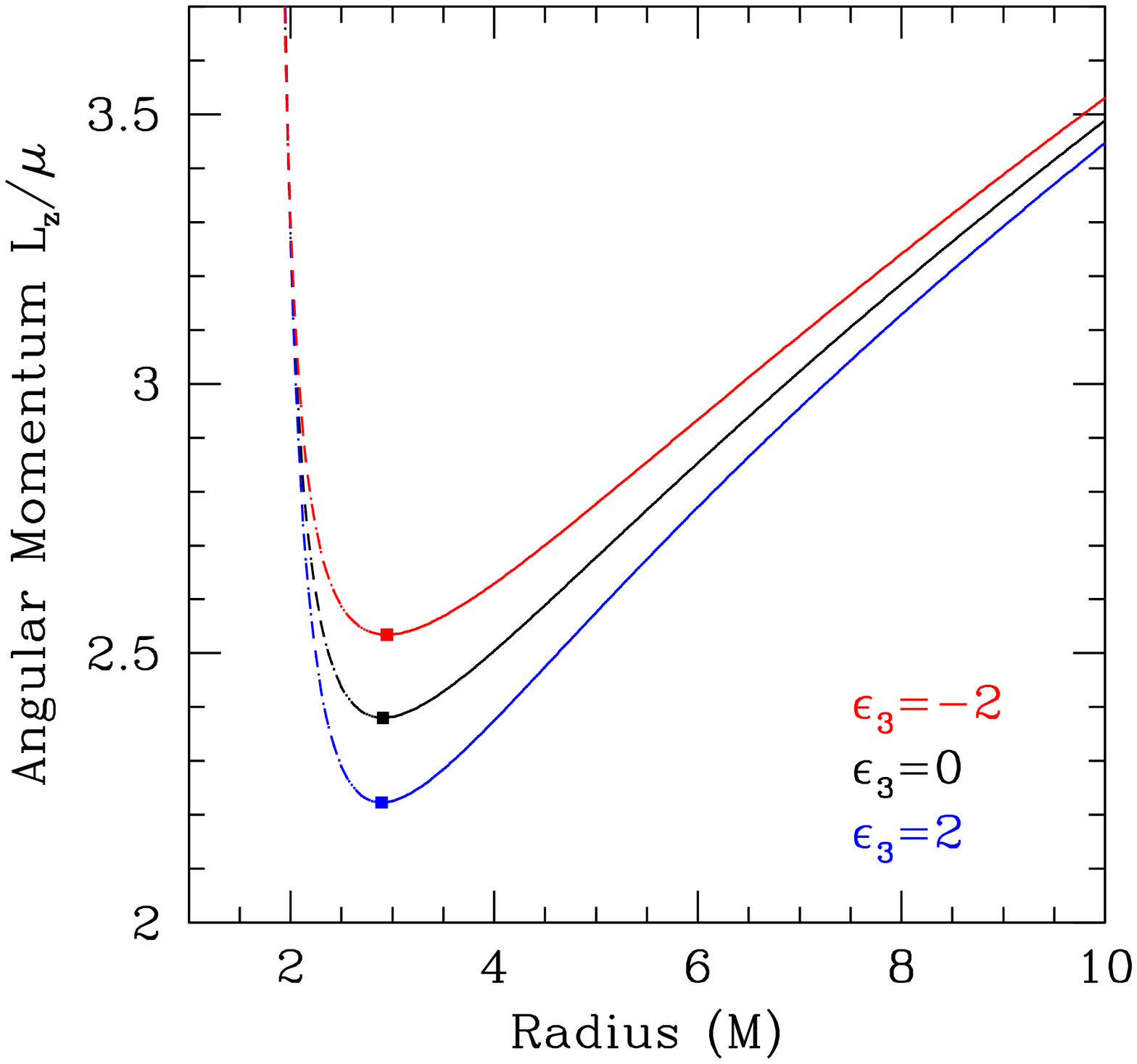,width=0.32\textwidth}
\psfig{figure=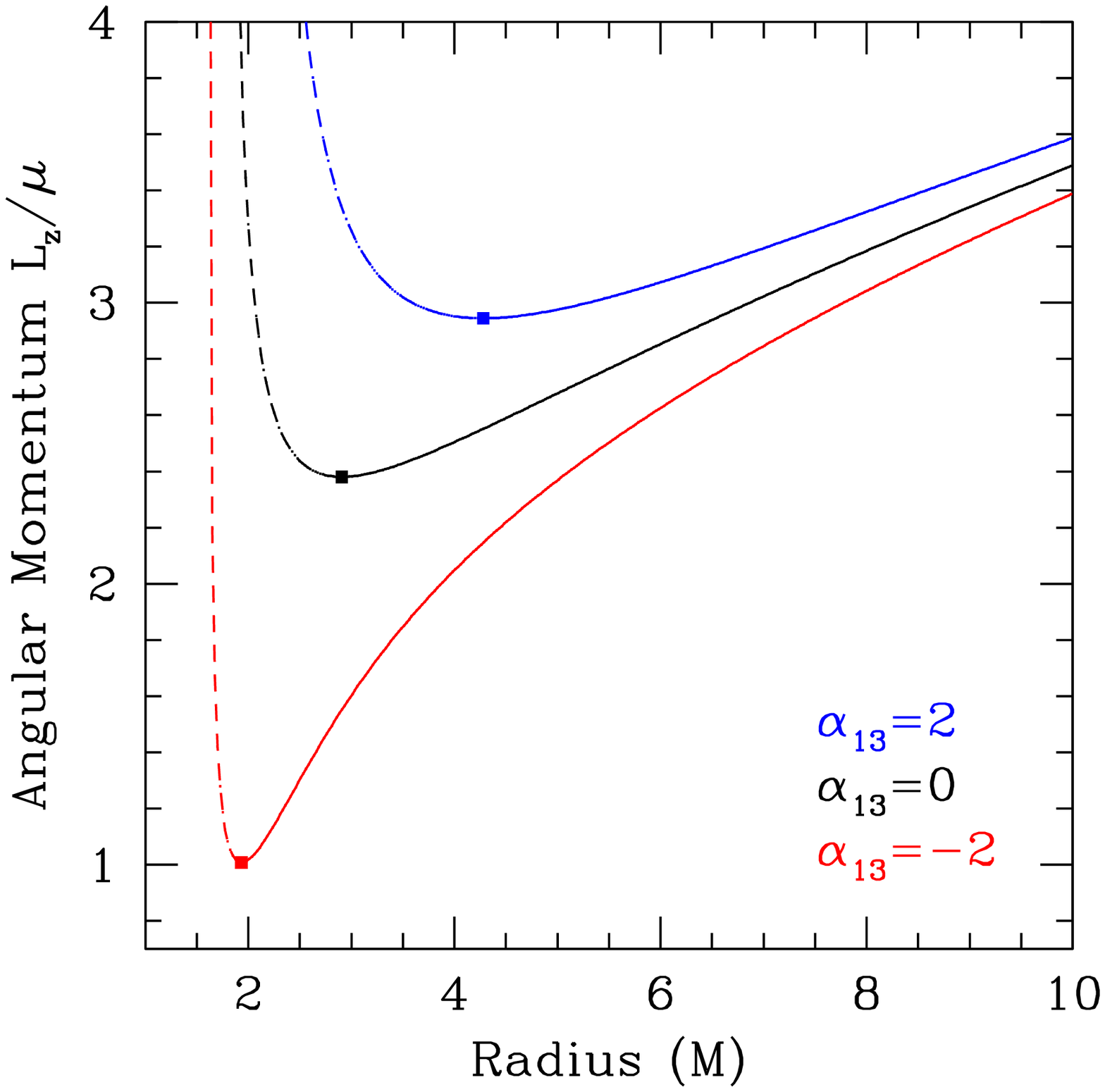,width=0.32\textwidth}
\psfig{figure=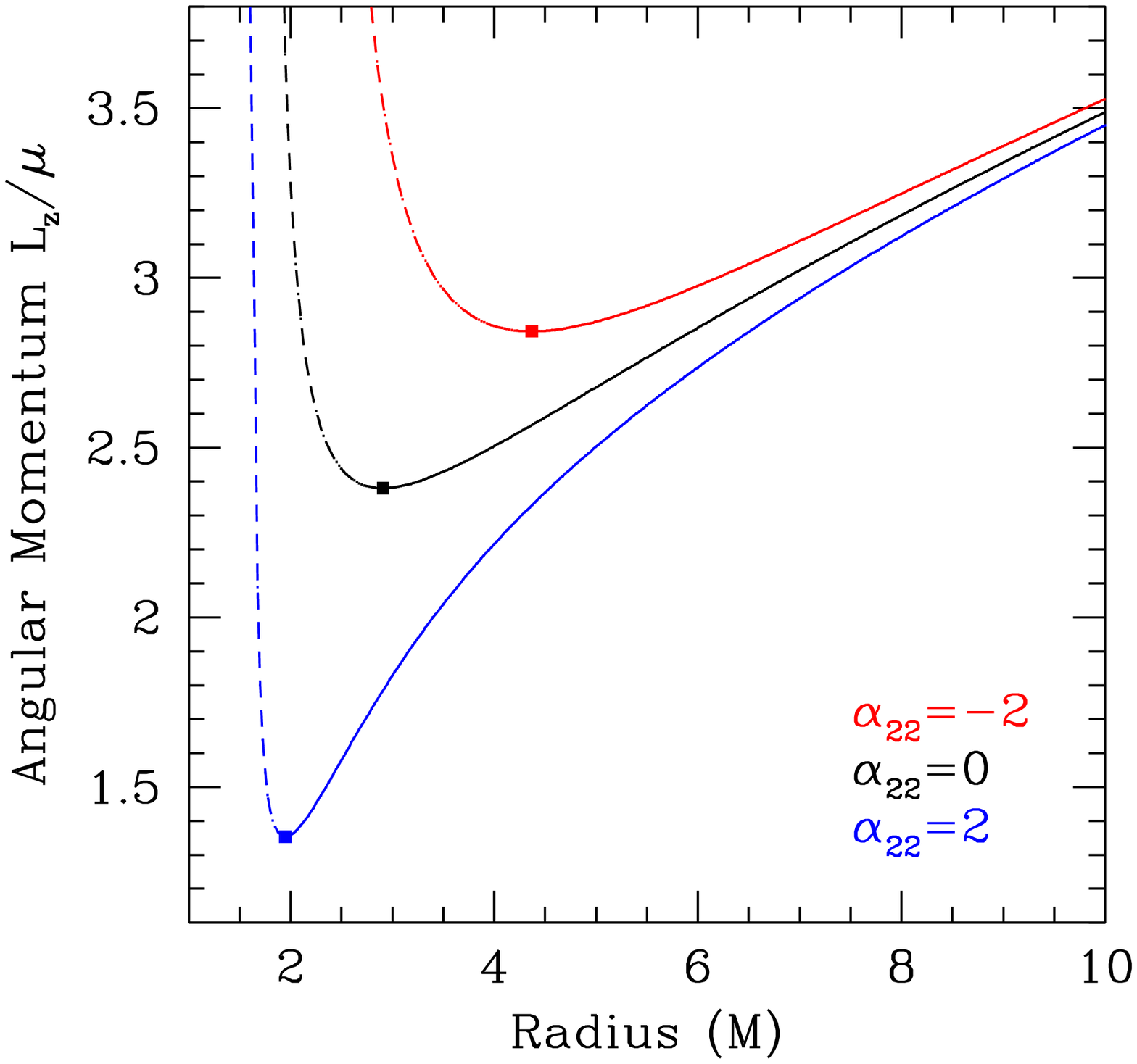,width=0.32\textwidth}
\end{center}
\caption{Dependence of the axial angular momentum $L_z$ about the black-hole spin axis of a particle with rest mass $\mu$ on a circular equatorial orbit around a black hole with spin $a=0.8M$ as observed at infinity on the deviation parameters $\epsilon_3$, $\alpha_{13}$, and $\alpha_{22}$. Each panel shows the axial angular momentum as a function of radius for different values of one deviation parameter while setting the other two equal to zero. At a given radius, the axial angular momentum increases for decreasing values of the parameters $\epsilon_3$ and $\alpha_{22}$ and for increasing values of the parameter $\alpha_{13}$. The dot denotes the location of the ISCO.}
\label{fig:lz}
\end{figure*}

\begin{figure*}[ht]
\begin{center}
\psfig{figure=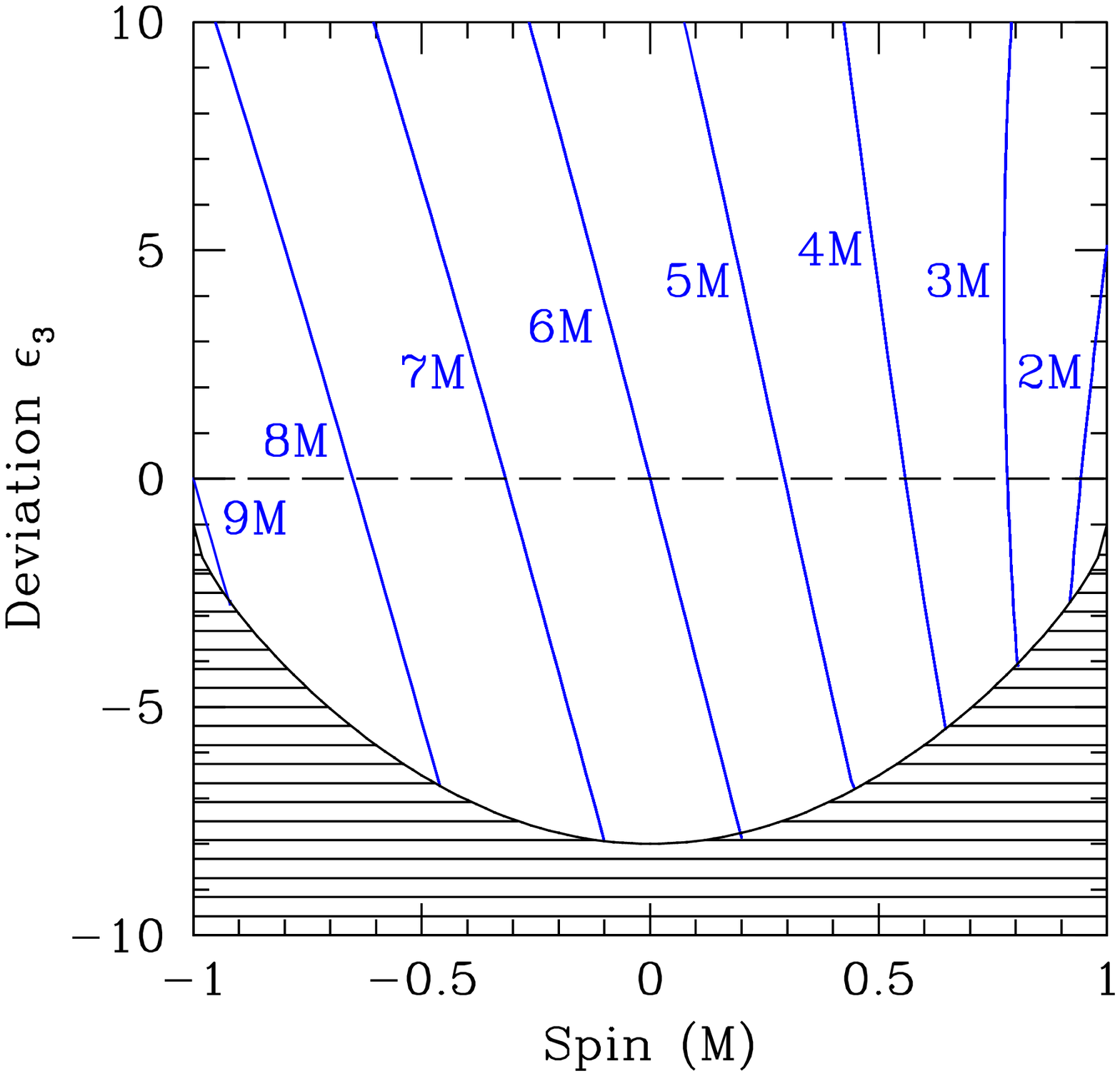,width=0.32\textwidth}
\psfig{figure=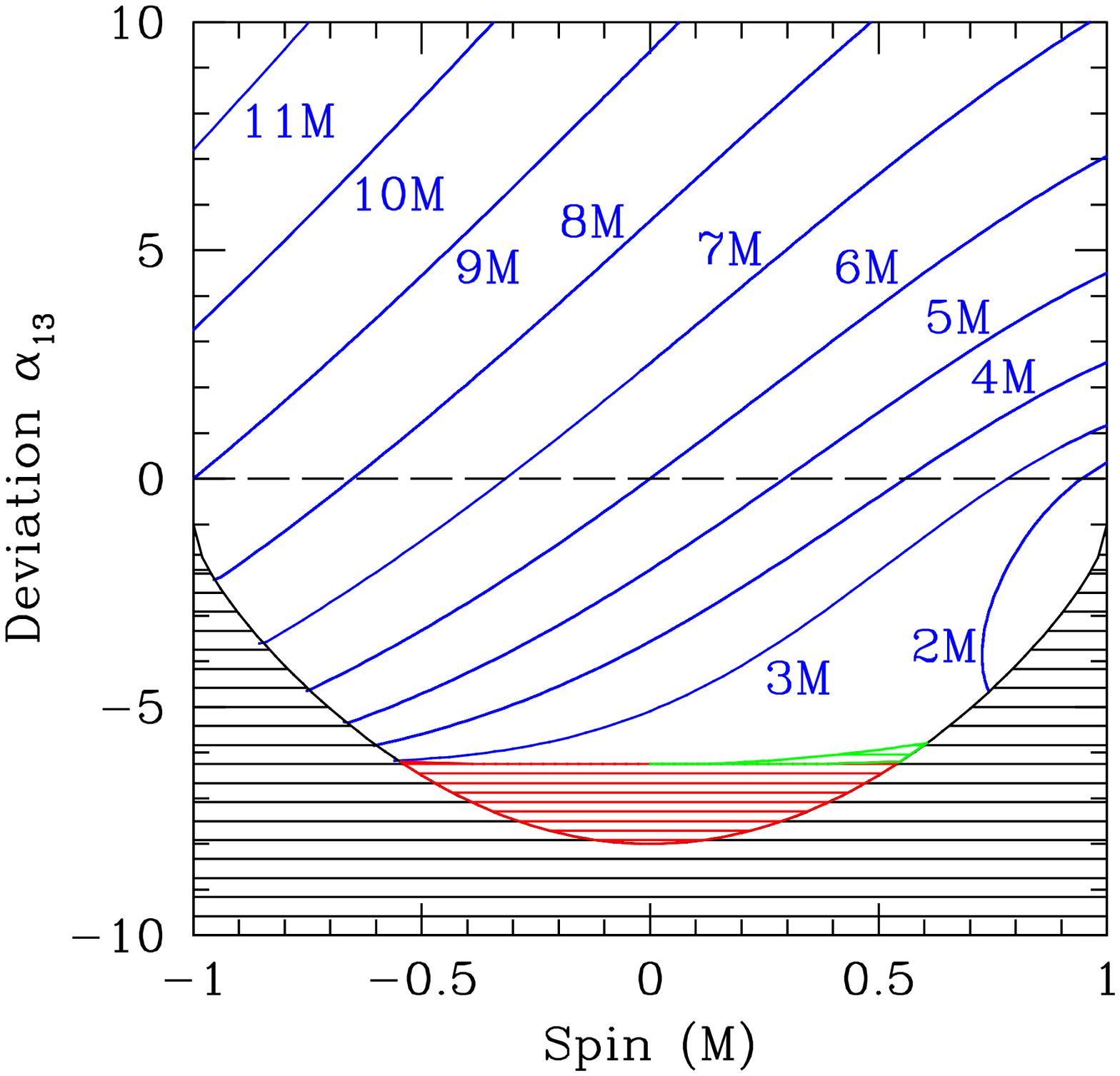,width=0.32\textwidth}
\psfig{figure=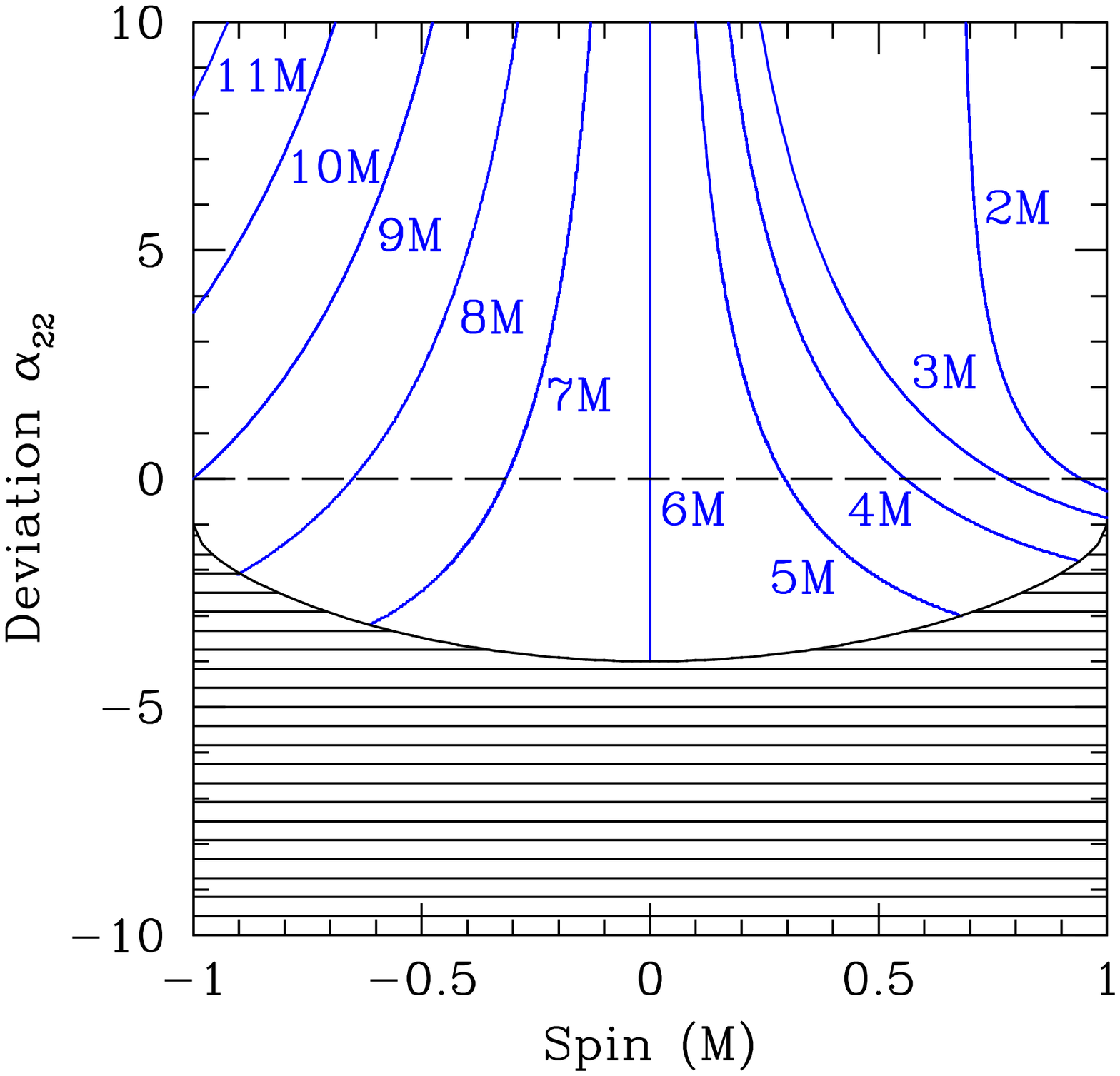,width=0.32\textwidth}
\end{center}
\caption{Dependence of the ISCO radius on the deviation parameters $\epsilon_3$, $\alpha_{13}$, and $\alpha_{22}$. Each panel shows contours of constant ISCO radius as a function of the spin and one deviation parameter while setting the other two equal to zero. At a fixed value of the spin, the location of the ISCO can either increase or decrease for increasing values of each deviation parameter and, in some cases, be practically independent of each deviation parameter. In the green shaded region of the central panel, the energy has two local minima and the ISCO is located at the outer radius where these minima occur. In the red shaded region of the central panel, circular equatorial orbits do not exist at radii $r\sim2.5M$ and the ISCO is located at the outer boundary of this radial interval. The black shaded region marks the excluded part of the parameter space.}
\label{fig:isco}
\end{figure*}

\begin{figure}[ht]
\begin{center}
\psfig{figure=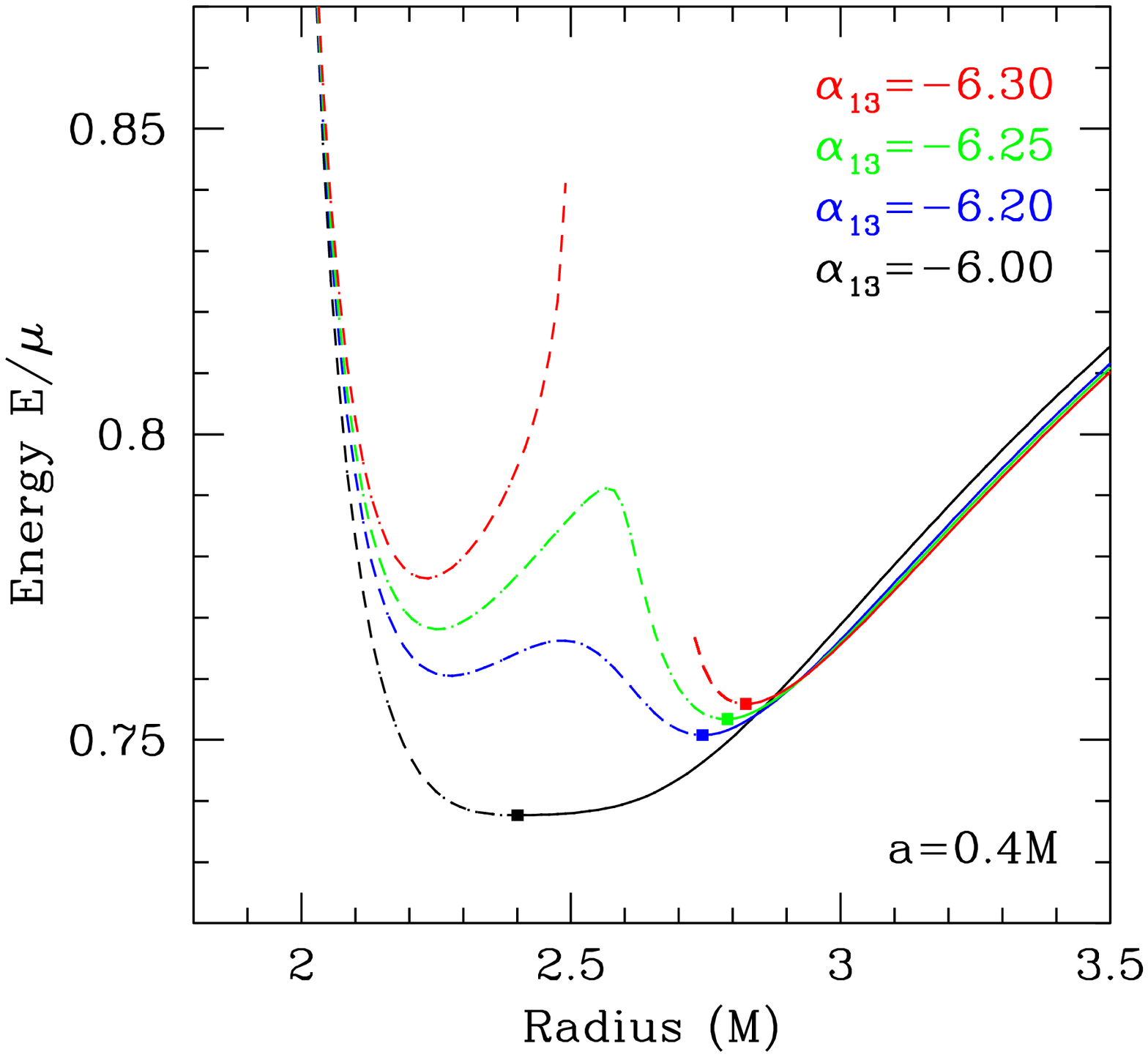,height=2.93in}
\psfig{figure=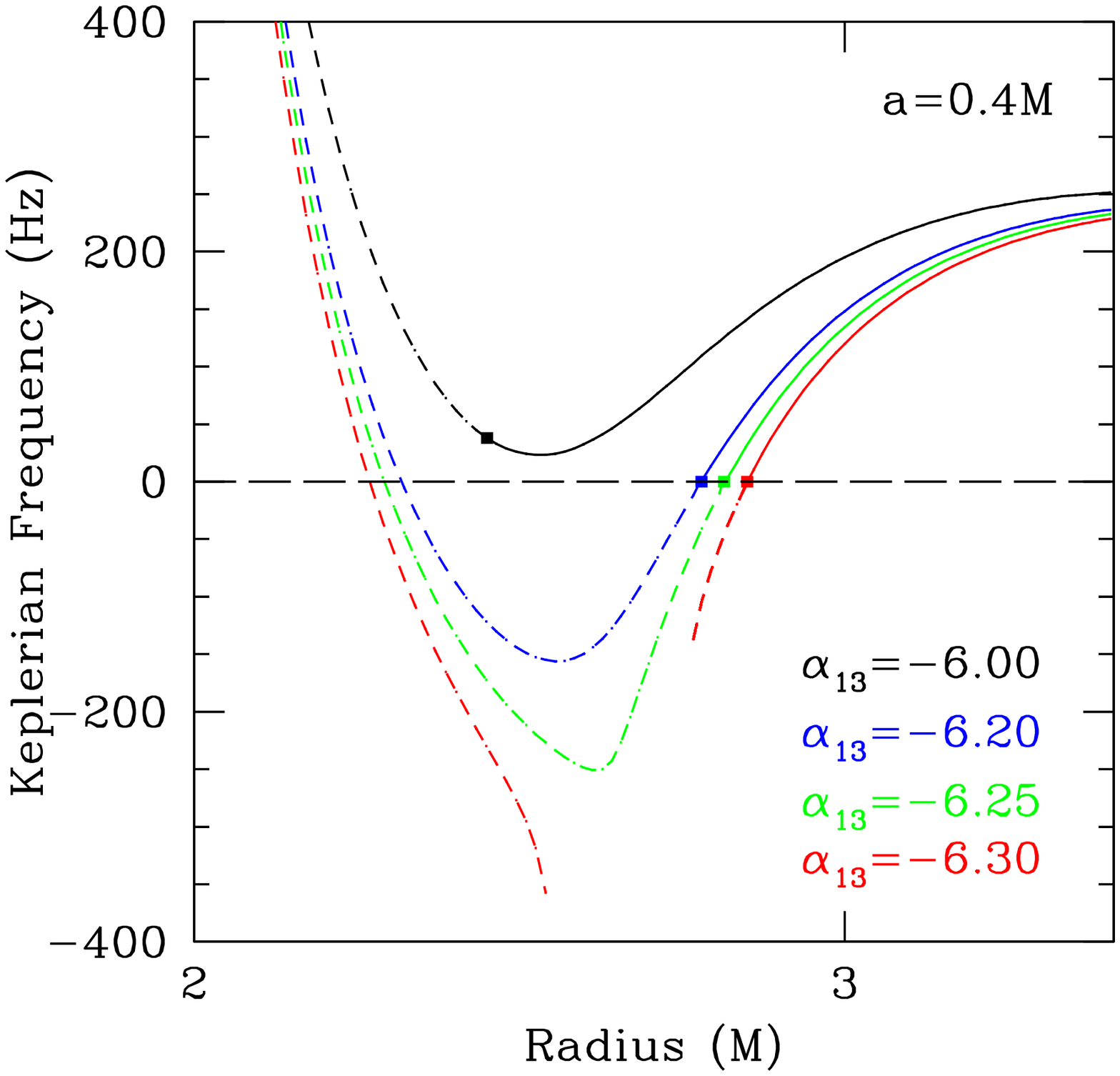,height=2.93in}
\end{center}
\caption{Energy (top) and Keplerian frequency (bottom) of a particle on a circular equatorial orbit around a black hole with spin $a=0.4M$ for several values of the parameter $\alpha_{13}$; all other deviation parameters are set to zero. For large negative values of the parameter $\alpha_{13}$, the energy has a second local minimum inside of the ISCO radius (labeled by a dot) with a higher value of the energy than at the ISCO. The Keplerian frequency decreases at radii just outside of the ISCO and vanishes approximately at the ISCO radius. For even more negative values, both the energy and the Keplerian frequency are imaginary at radii $r\sim2.5M$ and circular equatorial orbits no longer exist. In the latter case, the ISCO lies at the outer boundary of this radial interval.}
\label{fig:2iscos}
\end{figure}

In the metric designed in Ref.~\cite{JPmetric}, for extreme values of the deviation parameters and at certain radii very close to the ISCO, sign changes can occur in the expressions of the energy and axial angular momentum so that these quantities are smooth at these radii; see Ref.~\cite{OSCO} for a detailed discussion. In principle, a similar behavior in these expressions of the metric designed in this paper could arise, if the discriminant of the square root in the denominator of Eqs.~\eqref{eq:E} and \eqref{eq:Lz} has roots. Empirically, by evaluating these functions for different values of the deviation parameters, I find that such roots exist at least in the case of the lowest-order metric. However, the energy and axial angular momentum are still smooth at these roots and I have seen no evidence for discontinuities of the first derivatives of the energy or axial angular momentum outside of the ISCO. Should these nonetheless exist, then the opposite sign of the square roots in Eqs.~\eqref{eq:E} and \eqref{eq:Lz} has to be chosen at radii smaller than the root of the discriminant.

In Figs.~\ref{fig:en} and \ref{fig:lz}, I plot the energy and axial angular momentum versus the radius $r$ for different values of the lowest-order deviation parameters. As before, on varying one of these parameters, I set the other two equal to zero. At a given radius, both the energy and the axial angular momentum increase for decreasing values of the parameters $\epsilon_3$ and $\alpha_{22}$ and for increasing values of the parameter $\alpha_{13}$.

I calculate the location of the ISCO from the equation
\be
\frac{dE}{dr} = 0.
\ee
In Fig.~\ref{fig:isco}, I plot contours of constant ISCO radius versus the spin for different values of the deviation parameters. Again, on varying one of these parameters, I set the other two equal to zero.

The dependence of the ISCO on these parameters is more complex. As shown in Fig.~\ref{fig:isco}, at a fixed value of the spin, the location of the ISCO decreases for increasing values of the parameter $\epsilon_3$ for values of the spin $a\lesssim0.8M$. For values of the spin $a\gtrsim0.8M$, the ISCO radius increases for increasing values of the parameter $\epsilon_3$. At a spin $a\approx0.8M$, the ISCO is practically independent of the parameter $\epsilon_3$. The location of the ISCO increases for increasing values of the parameter $\alpha_{13}$ except for values of the spin $a\gtrsim0.7M$ and values of the parameter $\alpha_{13}$ which are very close to the boundary as defined in Eq.~\eqref{eq:alpha13cond}, where the ISCO becomes practically independent of the parameter $\alpha_{13}$. If the ISCO is located at a radius $r_{\rm ISCO}\lesssim3M$, it depends only weakly on the parameter $\alpha_{13}$.

For values of the parameter $\alpha_{13}\lesssim-5.7$, the energy $E$ can have two local minima. In this case, stable circular orbits exist only at and outside of the radius where the energy has a local maximum between its two local minima, because the radial epicyclic frequency is imaginary otherwise (see the discussion below), and the ISCO is located at the outer radius where the minima occur. For values of the parameter $\alpha_{13}\lesssim-6.2$, both the energy and Keplerian frequency become imaginary in a small range of radii $r\sim2.5M$ and the ISCO simply lies at the outer boundary of this radial interval even if the energy does not have a local minimum at this location. In certain cases, the Keplerian frequency vanishes approximately at the ISCO. I illustrate this behavior of the energy and Keplerian frequency in Fig.~\ref{fig:2iscos}.

Finally, the ISCO radius is independent of the parameter $\alpha_{22}$ if $a=0$ and depends only weakly on the parameter $\alpha_{22}$ if $|a|\sim0$ or if $r_{\rm ISCO}\lesssim3M$. If $a>0$, the location of the ISCO decreases for increasing values of the parameter $\alpha_{22}$, while the location of the ISCO increases for increasing values of the parameter $\alpha_{22}$ if $a<0$.

In order to derive expressions for the radial and vertical epicyclic frequencies $\Omega_r$ and $\Omega_\theta$, I write Eq.~\eqref{eq:Veff=0} in the forms
\ba
\frac{1}{2} \left( \frac{dr}{dt} \right)^2 &=& \frac{ V_{\rm eff} }{ g_{\rm rr}(p^{\rm t})^2 } \equiv V_{\rm eff}^{\rm r}, 
\label{eq:Veffr} \\
\frac{1}{2} \left( \frac{d\theta}{dt} \right)^2 &=& \frac{ V_{\rm eff} }{ g_{\rm \theta\theta}(p^{\rm t})^2 } \equiv V_{\rm eff}^{\rm \theta}.
\label{eq:Vefftheta}
\ea

Now I introduce small perturbations $\delta r$ and $\delta \theta$ and take the derivative of Eqs.~\eqref{eq:Veffr} and \eqref{eq:Vefftheta} with respect to the coordinate time, which yields the equations
\ba
\frac{d^2(\delta r)}{dt^2} &=& \frac{ d^2 V_{\rm eff}^{\rm r} }{ dr^2 } \delta r, \\
\frac{d^2(\delta \theta)}{dt^2} &=& \frac{ d^2 V_{\rm eff}^{\rm \theta} }{ d\theta^2 } \delta \theta.
\ea
From these expressions, I derive the radial and vertical epicyclic frequencies as
\ba
\Omega_{\rm r}^2 &=& -\frac{ d^2 V_{\rm eff}^{\rm r} }{dr^2}, \\
\label{kappa}
\Omega_{\rm \theta}^2 &=& -\frac{ d^2 V_{\rm eff}^{\rm \theta} }{ d\theta^2 },
\label{omegatheta}
\ea
where the second derivatives are evaluated at $r=r_0$. These expressions are lengthy and I do not write them here explicitly. Since the metric element $g_{rr}$ occurs only in Eq.~\eqref{eq:Veffr} and not in Eq.~\eqref{eq:Vefftheta}, the vertical epicyclic frequency is independent of the deviation function $A_5(r)$, while the radial epicyclic frequency depends on all four deviation functions.

\begin{figure*}[ht]
\begin{center}
\psfig{figure=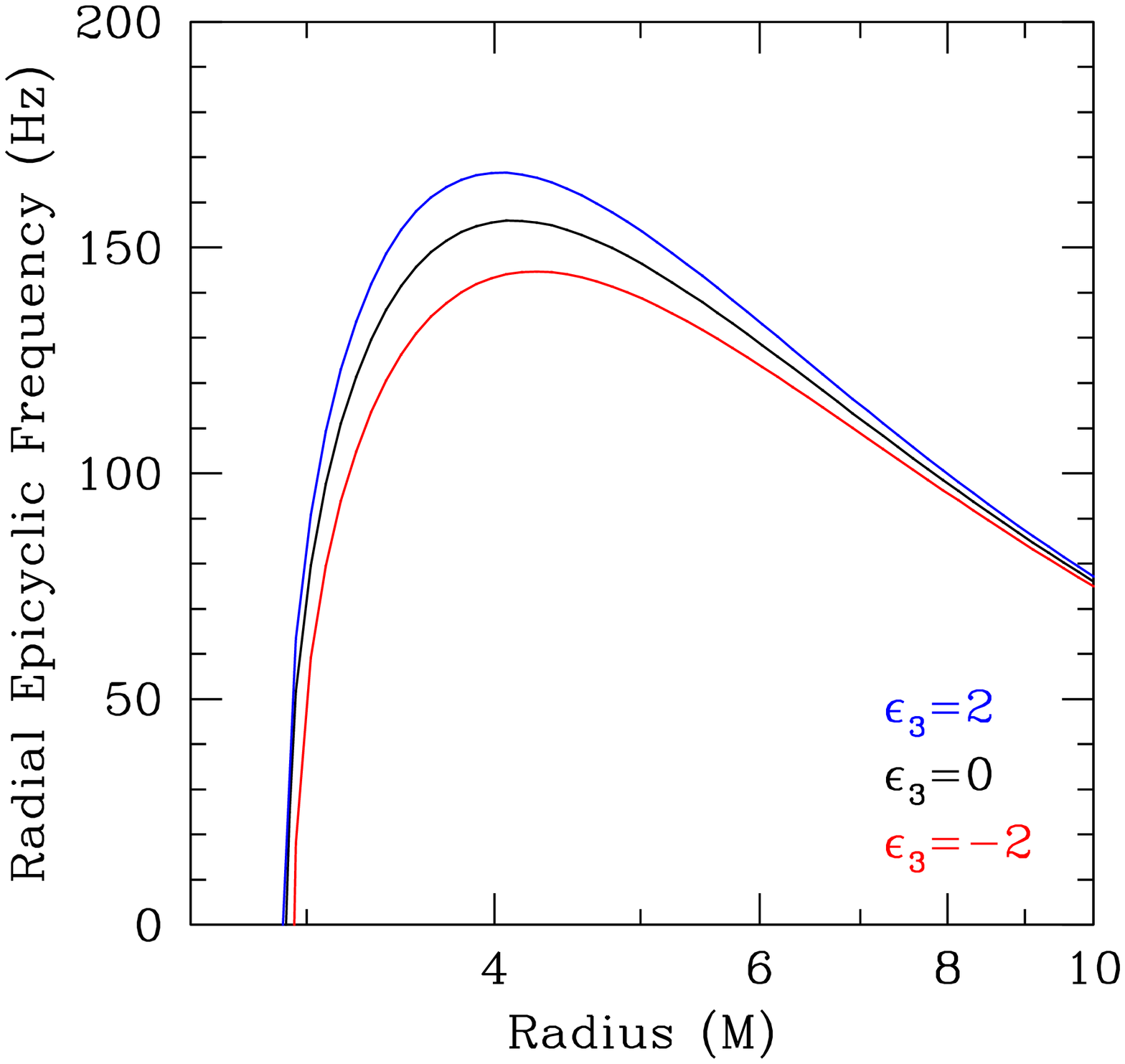,width=0.33\textwidth}
\psfig{figure=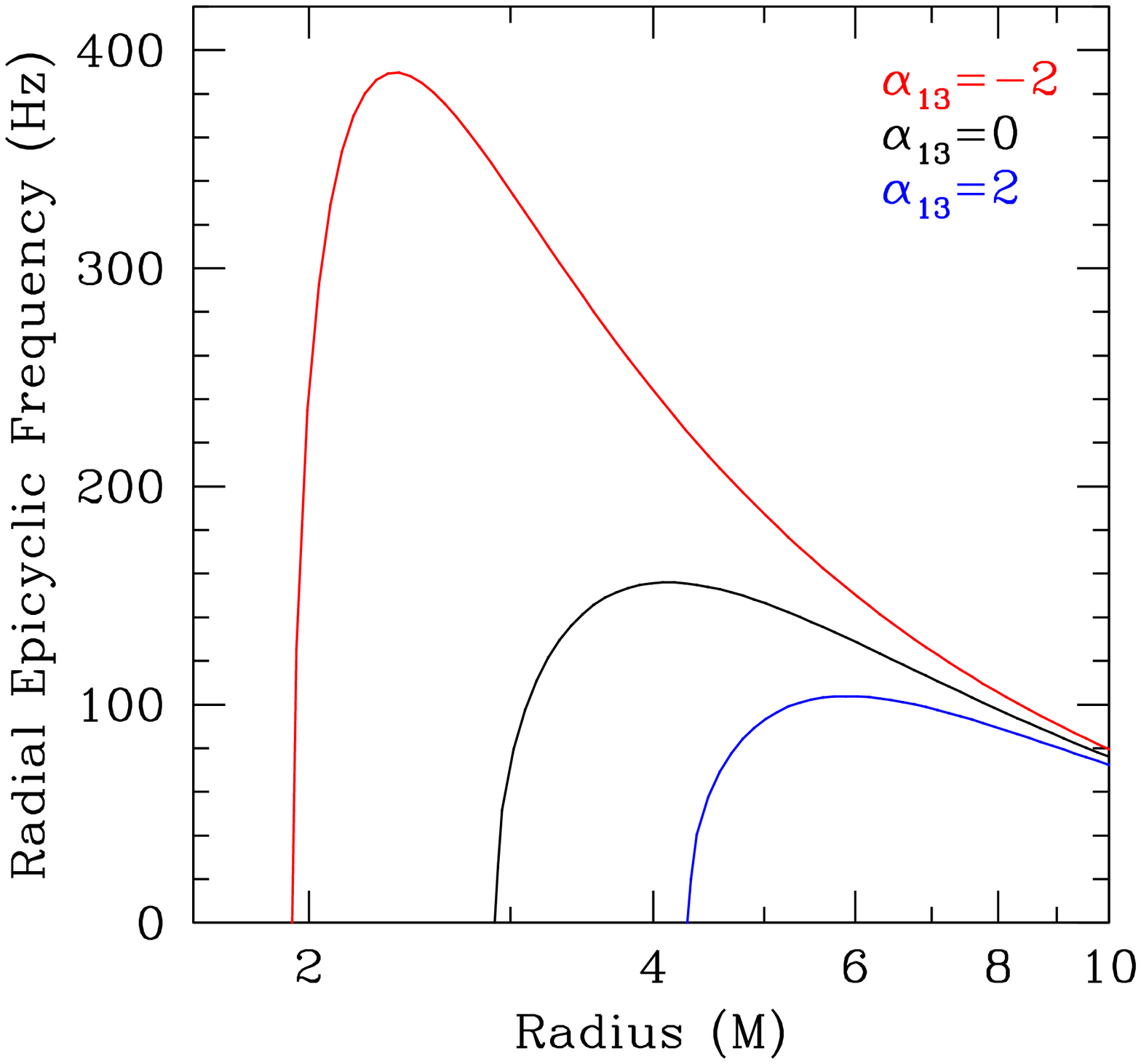,width=0.33\textwidth}
\psfig{figure=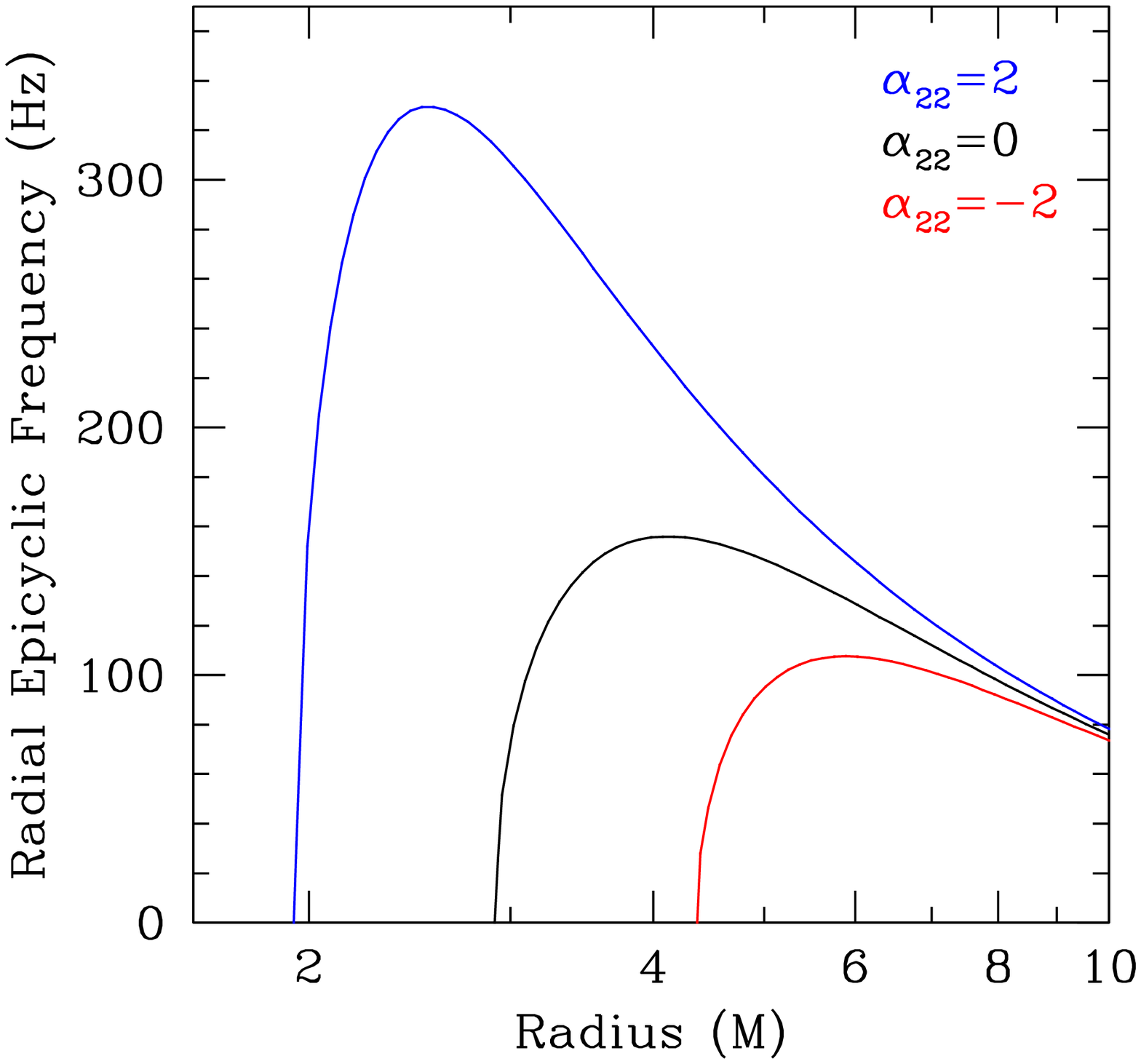,width=0.33\textwidth}
\psfig{figure=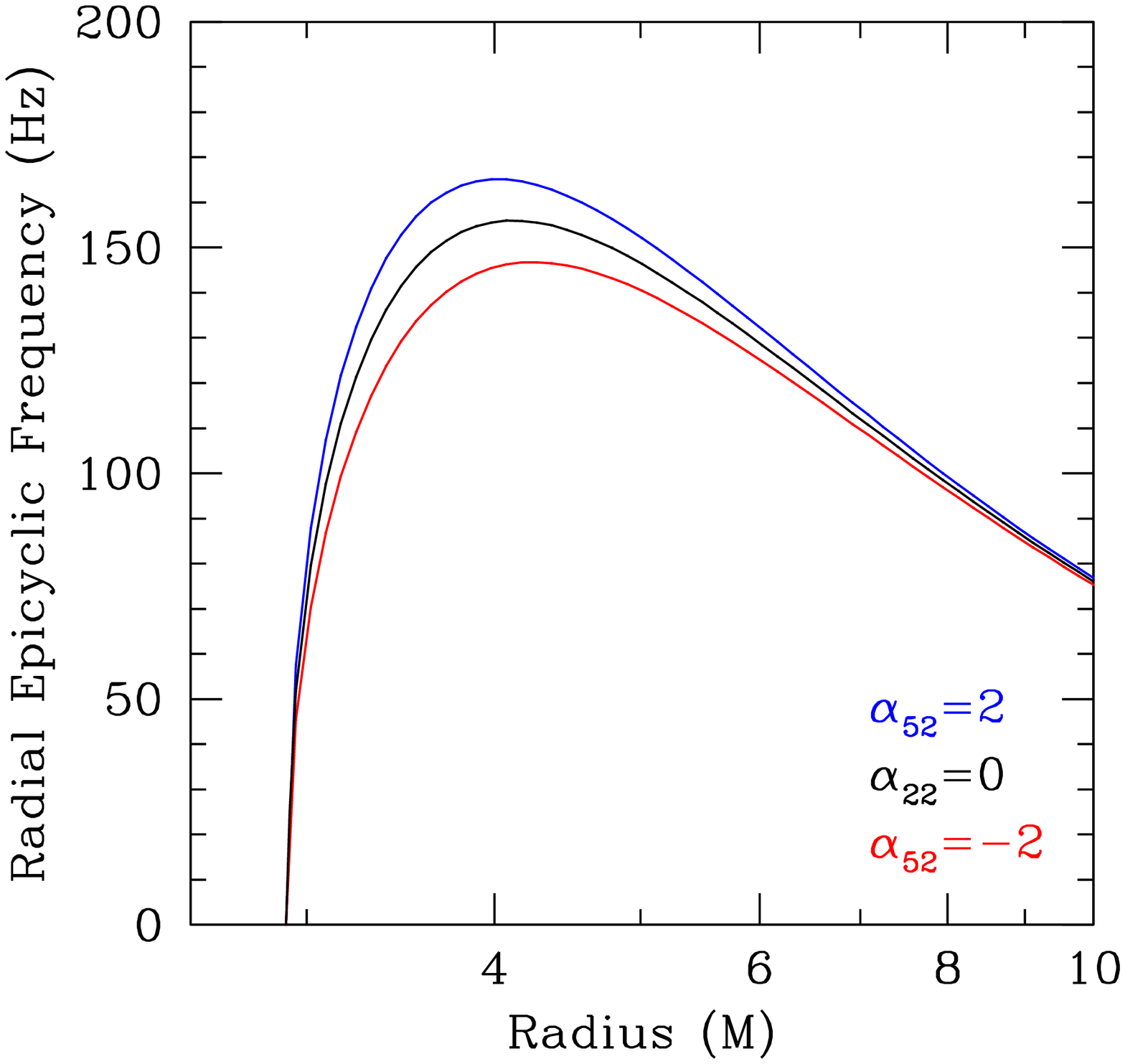,width=0.33\textwidth}
\end{center}
\caption{Dependence of the radial epicyclic frequency $\nu_r=c^3\Omega_r/2\pi GM$ on the deviation parameters $\epsilon_3$, $\alpha_{13}$, $\alpha_{22}$, and $\alpha_{52}$. Each panel shows the radial epicyclic frequency as a function of radius for a black hole with mass $M=10M_\odot$ and spin $a=0.8M$ for different values of one deviation parameter while setting the other two equal to zero. At a given radius, the radial epicyclic frequency increases for increasing values of the parameters $\epsilon_3$, $\alpha_{22}$, and $\alpha_{52}$ and for decreasing values of the parameter $\alpha_{13}$. The ISCO is located at the radius where the radial epicyclic frequency vanishes.}
\label{fig:kap}
\end{figure*}

\begin{figure*}[ht]
\begin{center}
\psfig{figure=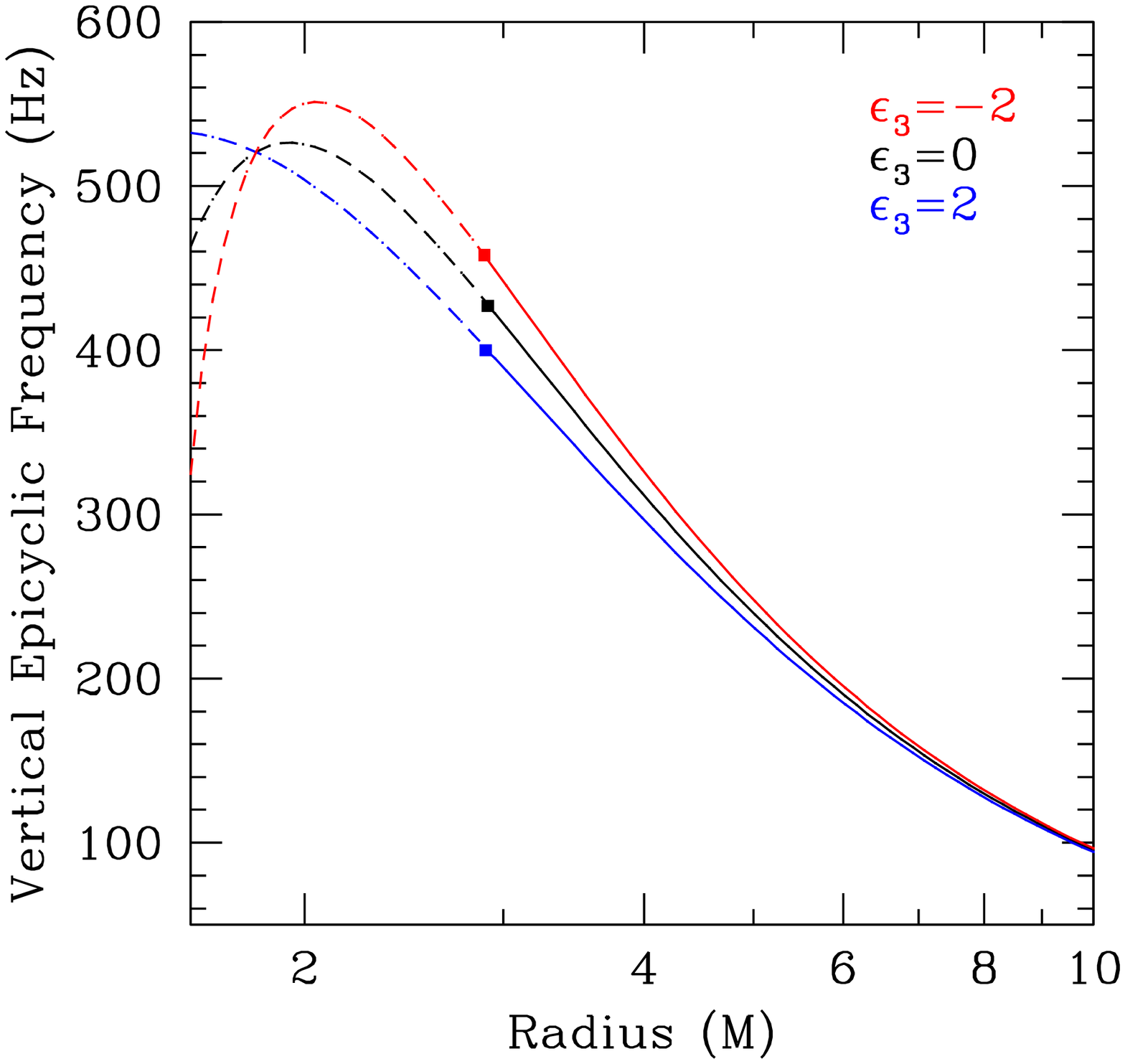,width=0.32\textwidth}
\psfig{figure=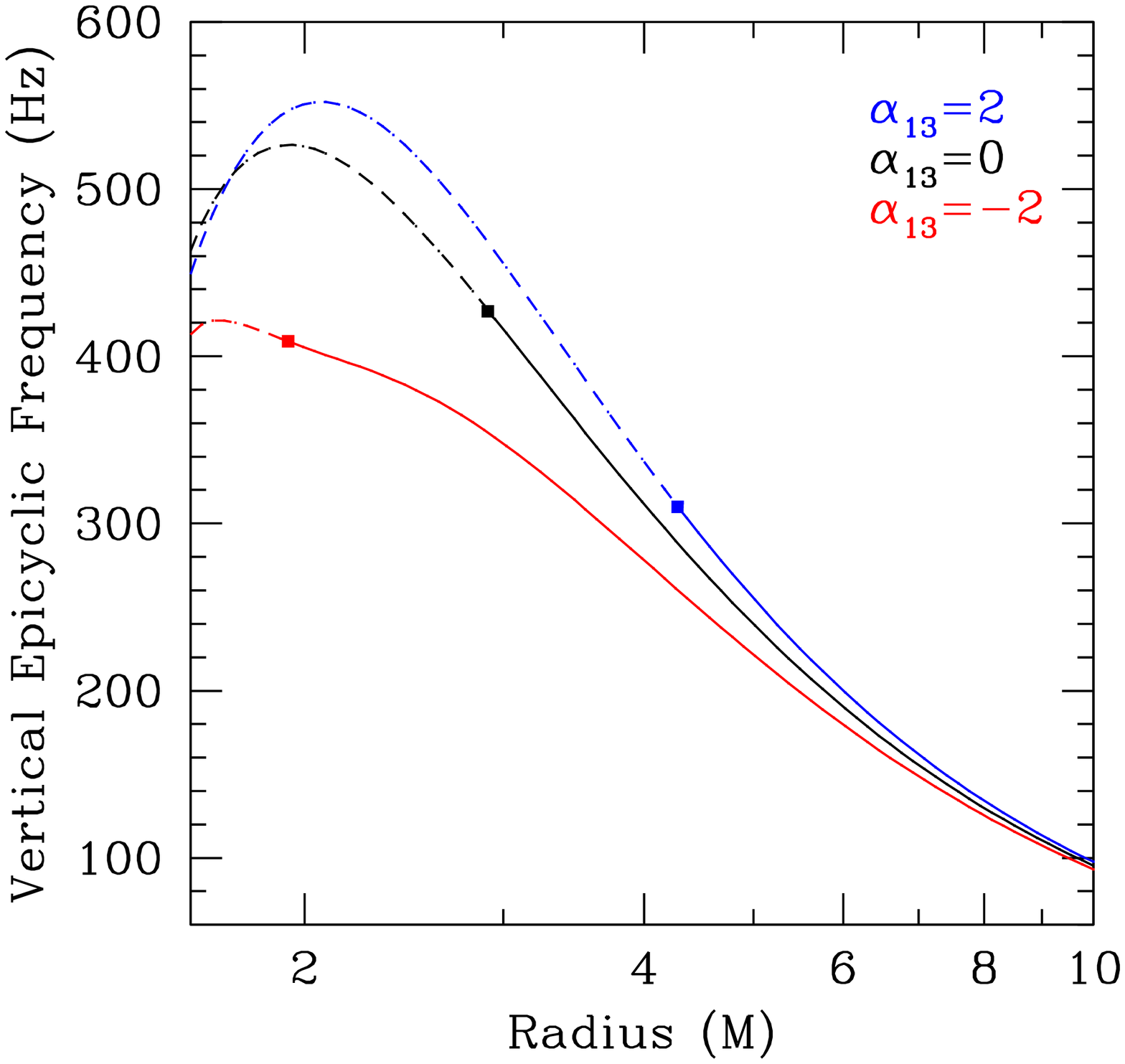,width=0.32\textwidth}
\psfig{figure=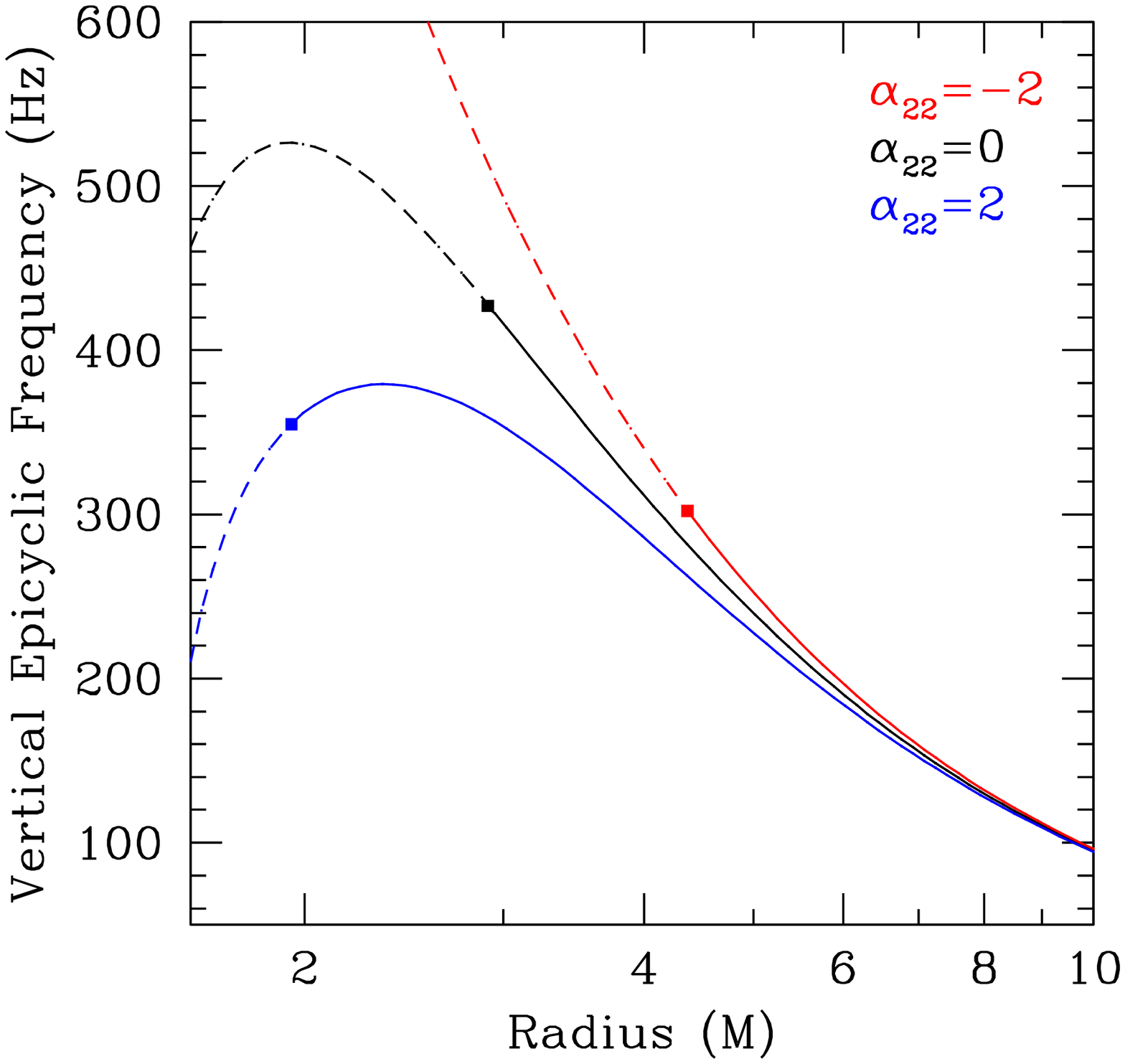,width=0.32\textwidth}
\end{center}
\caption{Dependence of the vertical epicyclic frequency $\nu_\theta=c^3\Omega_\theta/2\pi GM$ on the deviation parameters $\epsilon_3$, $\alpha_{13}$, and $\alpha_{22}$. Each panel shows the vertical epicyclic frequency $\nu_r$ as a function of radius for a black hole with mass $M=10M_\odot$ and spin $a=0.8M$ for different values of one deviation parameter while setting the other two equal to zero. At a given radius, the vertical epicyclic frequency increases for increasing values of the parameters $\epsilon_3$ and $\alpha_{22}$ and for decreasing values of the parameter $\alpha_{13}$.}
\label{fig:oth}
\end{figure*}

In Figs.~\ref{fig:kap} and \ref{fig:oth}, I plot the radial and epicyclic frequencies for a particle on a circular equatorial orbit around a black hole with mass $M=10M_\odot$ and spin $a=0.8M$ for different values of the deviation parameters. At a given radius, both the radial and the vertical epicyclic frequency increase for increasing values of the parameters $\epsilon_3$ and $\alpha_{22}$ and for decreasing values of the parameter $\alpha_{13}$. The radial epicyclic frequency also increases for increasing values of the parameter $\alpha_{52}$.

\begin{figure}[ht]
\begin{center}
\psfig{figure=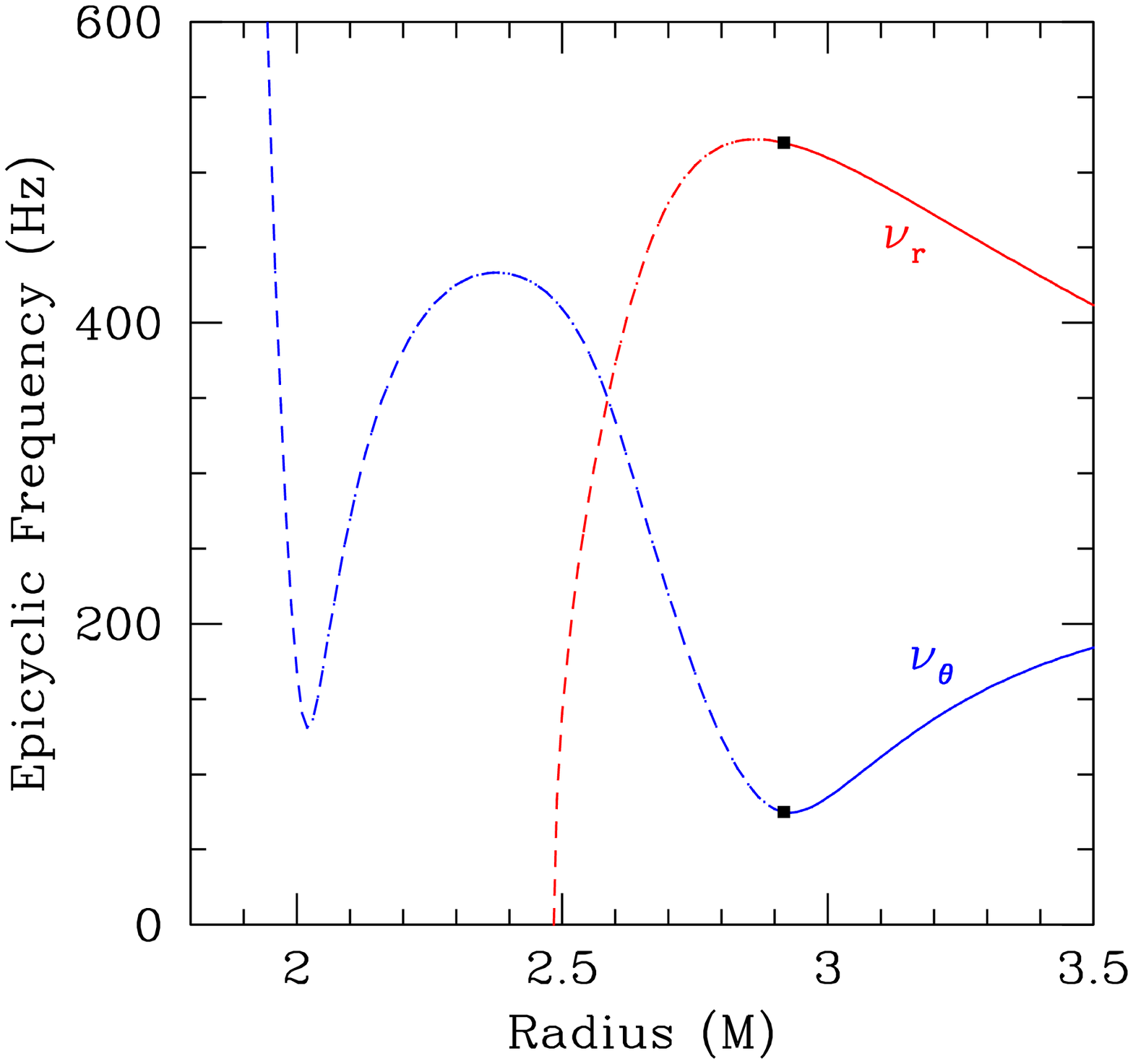,height=2.93in}
\end{center}
\caption{Radial and vertical epicyclic frequencies of a particle on a circular equatorial orbit around a black hole with mass $M=10M_\odot$ and spin $a=0.4M$ for a value of the parameter $\alpha_{13}=-6.2$; all other deviation parameters are set to zero. The black dots marks the location of the ISCO. The radial epicyclic frequency vanishes at the radius that corresponds to the local maximum of the energy $E$ (c.f., top panel in Fig.~\ref{fig:2iscos}). The vertical epicyclic frequency has a local minimum at the ISCO.}
\label{fig:2iscosfreqs}
\end{figure}

For all values of the deviation parameters in the allowed part of the parameter space as shown in Fig.~\ref{fig:parameterspace} up to values of at least $+10$, the radial epicyclic frequency always vanishes at some radius outside of the event horizon, which coincides with the ISCO unless the energy $E$ has more than one local minimum (c.f., the green and red shaded regions in Fig.~\ref{fig:isco}). Circular equatorial orbits are radially unstable inside of this radius and plunge into the black hole. For values of parameter $\alpha_{13}$ in the green shaded region of the central panel in Fig.~\ref{fig:isco}, the radial epicyclic frequency vanishes at the radius where the energy has a local maximum. In this case, the ISCO is located outside of this radius and the vertical epicyclic frequency has a minimum near the ISCO radius. I illustrate this property of the radial and vertical epicyclic frequencies in Fig.~\ref{fig:2iscosfreqs}, where I plot these frequencies for a black hole with mass $M=10M_\odot$ and spin $a=0.4M$ for a value of the parameter $\alpha_{13}=-6.2$.

\section{Kerr-Schild Form}
\label{sec:KSform}

In this section, I derive expressions for the principal null congruences of the metric given in Eq.~\eqref{eq:metric} and construct a transformation of the Boyer-Lindquist-like coordinates to Kerr-Schild-like coordinates. My derivation is similar to the corresponding calculations for the Kerr metric (c.f., e.g., Ref.~\cite{Chandrasekhar83}).

For null geodesics, it is convenient to introduce the parameters
\ba
\xi &\equiv& \frac{L_z}{E}, \\
\eta &\equiv& \frac{Q}{E^2},
\ea
so that the functions $R(r)$ and $\Theta(\theta)$ in Eqs.~\eqref{eq:R(r)} and \eqref{eq:Theta(theta)} can be written in the form (note that $\mu=0$ and $A_3(\theta)=A_4(\theta)=A_6(\theta)=1)$
\ba
\frac{R(r)}{E^2} &=& [(r^2+a^2)A_1(r)-aA_2(r)\xi]^2 \nn \\
&& -\Delta[ (\xi-a)^2 + \eta ],
\label{eq:Rng} \\
\frac{\Theta(\theta)}{E^2} &=& \eta + (\xi-a)^2 - \frac{1}{\sin^2\theta}(\xi-a\sin^2\theta)^2.
\label{eq:Thetang}
\ea

Since from Eq.~\eqref{eq:S_theta} $\Theta(\theta)\geq0$,
\be
\eta + (\xi-a)^2 \geq 0,
\ee
where equality holds if and only if
\ba
\theta &=& \theta_0 = {\rm const.}, \\
\xi &=& a\sin^2\theta.
\label{eq:xi}
\ea
In this case,
\be
\eta = -a^2\cos^4\theta_0.
\label{eq:eta}
\ee
These expressions for the parameters $\xi$ and $\eta$ are identical to the ones for the Kerr metric.

Using the equations of motion, Eqs.~\eqref{eq:eom_t}--\eqref{eq:eom_phi}, and Eqs.~\eqref{eq:xi} and \eqref{eq:eta}, I obtain the photon equations of motion
\ba
\frac{dt}{d\lambda} &\equiv& l^t E = \frac{(r^2+a^2)A_1(r)}{\Delta} \nn \\
&& \frac{(r^2+a^2)A_1(r)-a^2A_2(r)\sin^2\theta_0}{\tilde{\Sigma}}E, \nn \\
\frac{dr}{d\lambda} &\equiv& \pm l^r E = \pm \sqrt{A_5(r)} \nn \\
&& \frac{(r^2+a^2)A_1(r)-a^2A_2(r)\sin^2\theta_0}{\tilde{\Sigma}} E, \nn \\
\frac{d\theta}{d\lambda} &=& 0, \nn \\
\frac{d\phi}{d\lambda} &\equiv& l^\phi E = \frac{aA_2(r)}{\Delta} \nn \\
&& \frac{(r^2+a^2)A_1(r)-a^2A_2(r)\sin^2\theta_0}{\tilde{\Sigma}}E,
\label{eq:nulleqs}
\ea
where $\lambda$ is an affine parameter.

Accordingly, setting $E=1$, the principal null directions are given by the vectors
\be
l^\alpha_\pm = \left(l^t,\pm l^r,0,l^\phi\right).
\label{eq:png}
\ee

In order to remove the coordinate singularity of the metric in Eq.~\eqref{eq:metric} written in Boyer-Lindquist-like coordinates located at the event horizon $r_+$, I perform a transformation to Kerr-Schild-like coordinates, which can be defined using either the outgoing ($+l^r$) or ingoing ($-l^r$)  principal null direction in Eq.~\eqref{eq:png}. Defining the transformations
\ba
dt_{\rm KS} &\equiv& dt_{\rm BL}\mp\frac{l^t}{l^r} dr_{\rm BL}, \nn \\
dr_{\rm KS} &\equiv& \pm\frac{1}{l^r}dr_{\rm BL}, \nn \\
d\theta_{\rm KS} &\equiv& d\theta_{\rm BL}, \nn \\
d\phi_{\rm KS} &\equiv& d\phi_{\rm BL} \mp \frac{l^\phi}{l^r} dr_{\rm BL},
\label{eq:trafoBLKSgen}
\ea
the transformed principal null vectors (setting $E=1$) are given by the expression
\be
l^\alpha_\pm = (0,\pm1,0,0),
\label{eq:pngKS}
\ee
which is the form of the principal null directions of the Kerr metric in standard Kerr-Schild coordinates. Here, the upper/lower sign refers to the transformation in Eq.~\eqref{eq:trafoBLKSgen} with the upper/lower signs and the subscripts ``KS'' and ``BL'' stand for ``Kerr-Schild-like'' and ``Boyer-Lindquist-like'', respectively. Note, however, that the transformation of the radius makes this transformation rather cumbersome in practice and it may not be possible to perform it explicitly.

For this reason, I define an alternative transformation to Kerr-Schild-like coordinates, which drops the transformation of the radius in the transformation in Eq.~\eqref{eq:trafoBLKSgen} and which is given by the relations
\ba
dt_{\rm KS} &=& dt_{\rm BL}-\frac{2Mr A_1(r)}{\Delta \sqrt{A_5(r)}} dr_{\rm BL}, \nn \\
dr_{\rm KS} &=& dr_{\rm BL}, \nn \\
d\theta_{\rm KS} &=& d\theta_{\rm BL}, \nn \\
d\phi_{\rm KS} &=& d\phi_{\rm BL} - \frac{a A_2(r)}{\Delta \sqrt{A_5(r)}} dr_{\rm BL}.
\label{eq:trafoBLKS}
\ea
In these expressions, I used the definitions of the null vector in Eq.~\eqref{eq:png} given by Eq.~\eqref{eq:nulleqs} with a slight modification of the component $l^t$, where I replaced the factor $(r^2+a^2)/\Delta$ by $2Mr/\Delta$ as in McKinney and Gammie \cite{McKG04}. A transformation to Kerr-Schild-like coordinates can be defined for both of these choice.

The metric in the above Kerr-Schild-like coordinates is then given by the elements
\ba
g_{tt}^{\rm KS} &=& -\frac{\tilde{\Sigma}[\Delta-a^2A_2(r)^2\sin^2\theta]}{F}, \nn \\
g_{tr}^{\rm KS} &=& \frac{\tilde{\Sigma}}{\sqrt{A_5(r)}F} \{A_1(r)[2Mr+a^2A_2(r)^2\sin^2\theta] \nn \\
&& - a^2A_2(r)\sin^2\theta\}, \nn \\
g_{t\phi}^{\rm KS} &=& -\frac{a\tilde{\Sigma}[(r^2+a^2)A_1(r)A_2(r)-\Delta]\sin^2\theta}{F}, \nn \\
g_{rr}^{\rm KS} &=& \frac{\tilde{\Sigma}A_1(r)}{A_5(r)F} \{ A_1(r)[\Delta+4Mr+a^2A_2(r)^2\sin^2\theta] \nn \\
&& - 2a^2A_2(r)\sin^2\theta \}, \nn \\
g_{r\phi}^{\rm KS} &=& -\frac{a\tilde{\Sigma}\sin^2\theta }{ \sqrt{A_5(r)}F } [(r^2+a^2)A_1(r)^2A_2(r) \nn \\
&& +2MrA_1(r)-a^2A_2(r)\sin^2\theta], \nn \\
g_{\theta\theta}^{\rm KS} &=& \tilde{\Sigma}, \nn \\
g_{\phi\phi}^{\rm KS} &=& \frac{\tilde{\Sigma} \left[(r^2+a^2)^2A_1(r)^2-a^2\Delta\sin^2\theta\right]\sin^2\theta }{ F },
\label{eq:metric_KS}
\ea
where
\be
F \equiv \left[ (r^2+a^2)A_1(r)-a^2A_2(r)\sin^2\theta \right]^2.
\ee
In the Kerr case, $A_1(r)=A_2(r)=A_5(r)=1$, $f(r)=0$, and this metric reduces to the Kerr-Schild metric in the form given in Eq.~(4) of McKinney and Gammie \cite{McKG04}.

From the metric elements in Eq.~\eqref{eq:metric_KS} it is clear that the metric in Kerr-Schild-like coordinates no longer has a coordinate singularity at the radius $r_+$ given by Eq.~\eqref{eq:Kerrhor}. Evaluating the $(r,r)$ and $(\theta,\theta)$ elements of the contravariant metric and inserting them into the horizon equation \eqref{eq:hor_master}, I obtain an equation that is identical to Eq.~\eqref{eq:genmetrichor}. Therefore, the metric in Kerr-Schild-like coordinates still harbors an event horizon at the radius $r_+$.

\begin{figure}[ht]
\begin{center}
\psfig{figure=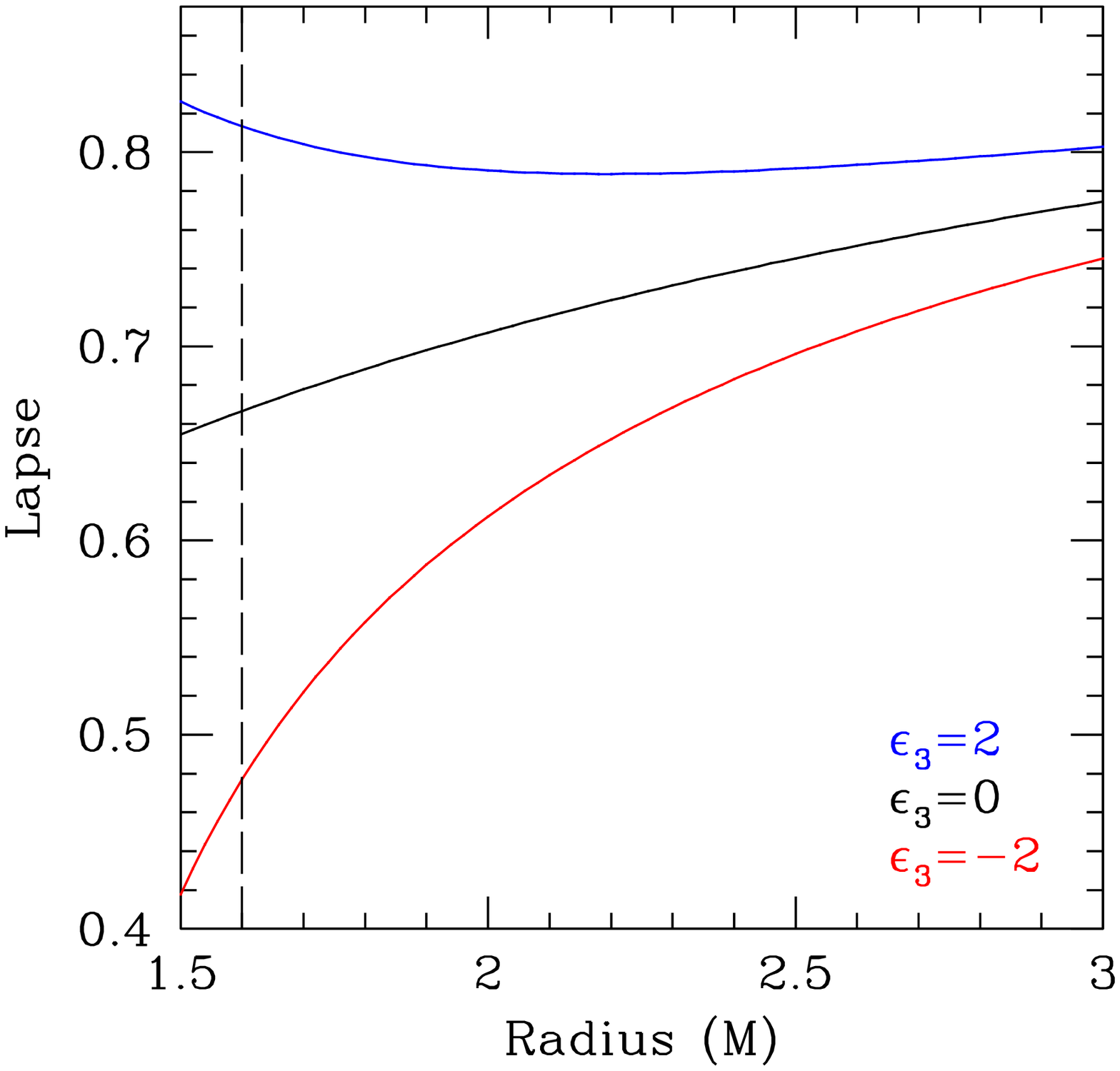,height=2.93in}
\psfig{figure=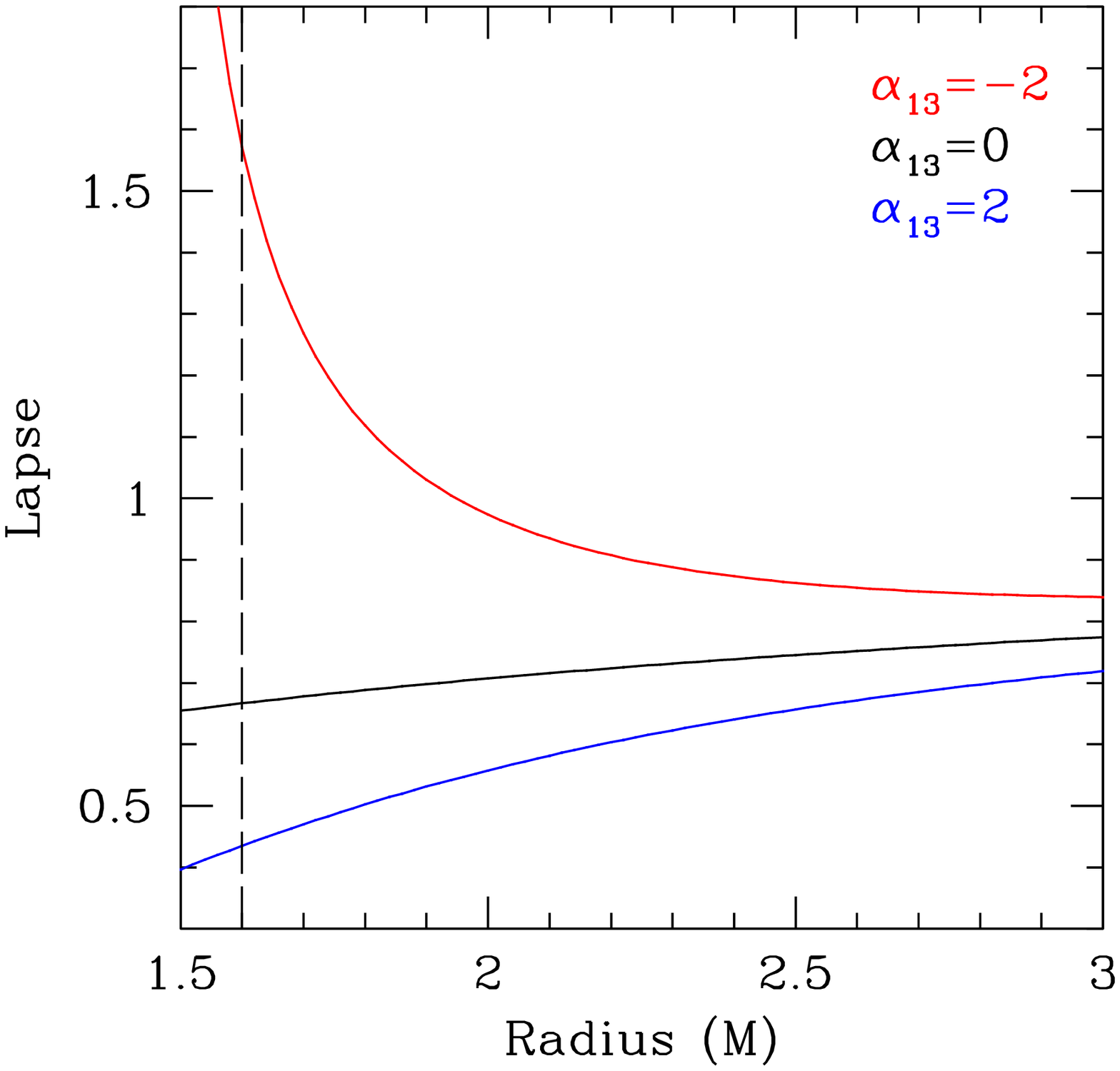,height=2.93in}
\end{center}
\caption{Lapse versus radius for a black hole with mass $M$ and spin $a=0.8M$ in the equatorial plane for different values of the parameters (top) $\epsilon_3$ and (bottom) $\alpha_{13}$. In each panel, only one parameter is varied, while the other one is set to zero. The lapse is positive across the radius that marks the location of the event horizon (vertical dashed line).}
\label{fig:lapse}
\end{figure}

The Kerr-Schild-like form of the metric in Eq.~\eqref{eq:metric_KS} is important for fully relativistic magnetohydrodynamic simulations of accretion flows in this metric, because these typically require the presence of an event horizon without any coordinate singularities at the horizon or in its vicinity both inside or outside of the horizon (see, e.g, Ref.~\cite{McKG04}). Otherwise, matter in the simulation might get trapped at the horizon leading to an unphysical matter accumulation. In particular, the lapse
\be
N\equiv\frac{1}{\sqrt{-g^{tt}_{\rm KS}}}
\ee
has to be positive across the event horizon.

The lapse of the lowest-order metric in Kerr-Schild-like form is given by the expression
\be
N = \sqrt{ \frac{ r^5(r^3+\epsilon_3M^3+a^2r\cos^2\theta) }{ (r^2+2Mr+a^2)(r^3+\alpha_{13}M^3)^2 - a^2r^6\sin^2\theta } },
\ee
which is indeed positive across the horizon in the allowed part of the parameter space [recall Eq.~\eqref{eq:Sigmatilde_cond}]. In Fig.~\ref{fig:lapse}, I plot the lapse for a black hole with mass $M$ and spin $a=0.8M$ in the equatorial plane for different values of the parameters $\epsilon_3$ and $\alpha_{13}$.

\begin{figure}[ht]
\begin{center}
\psfig{figure=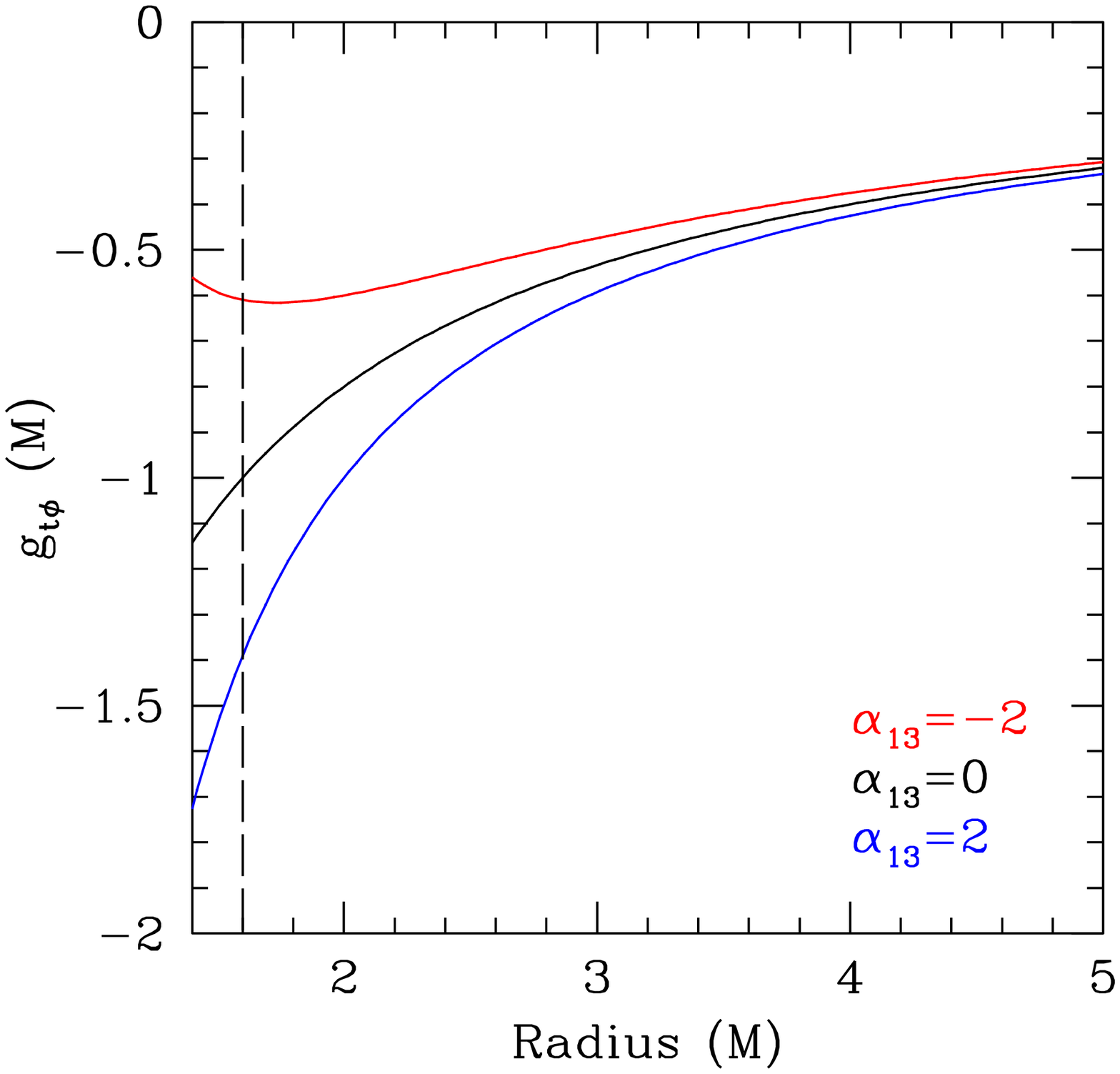,height=2.93in}
\psfig{figure=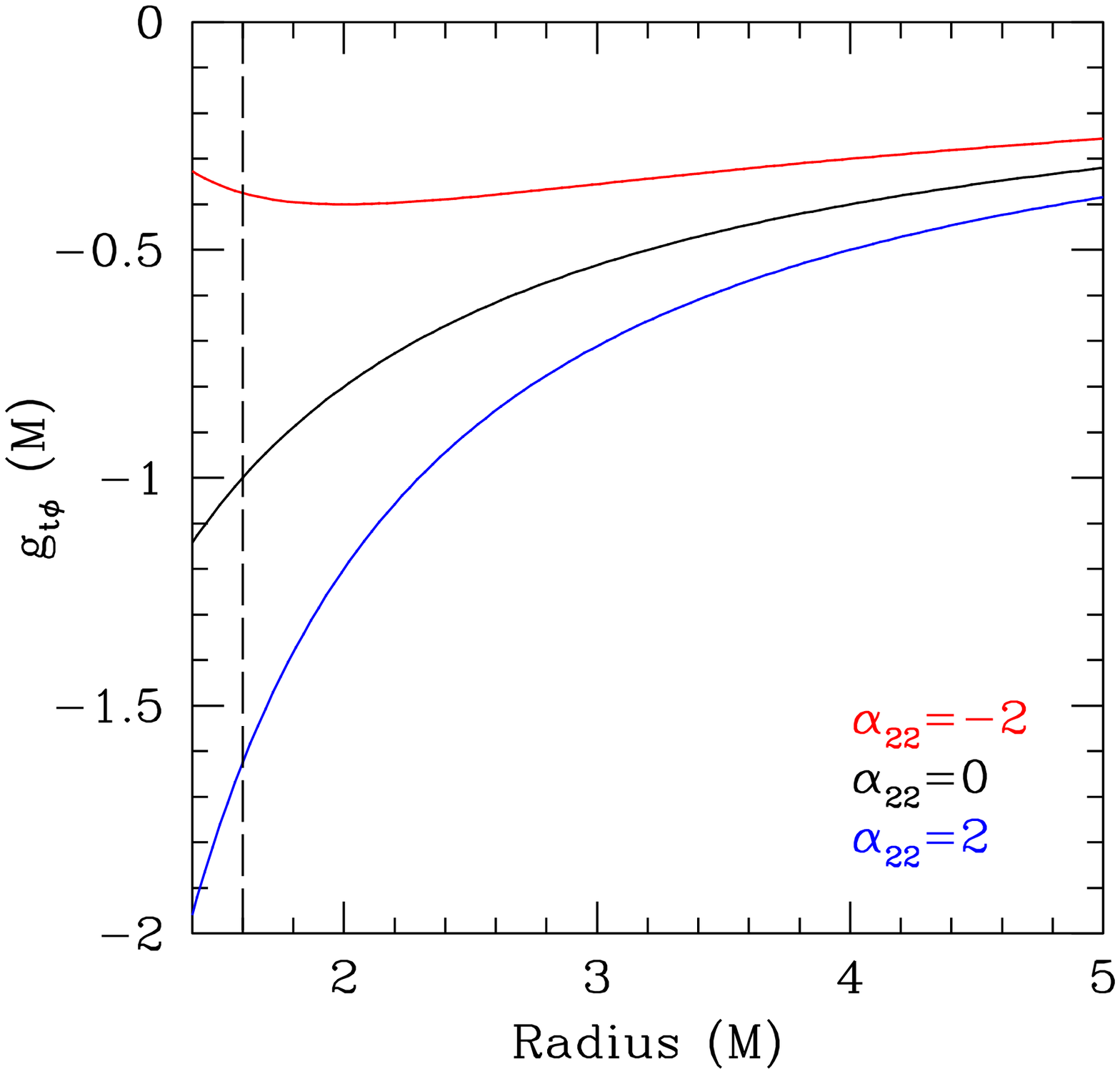,height=2.93in}
\end{center}
\caption{Element $g_{t\phi}^{\rm KS}$ of the lowest-order metric to ${\cal O}(1/r^3)$ versus radius for a black hole with mass $M$ and spin $a=0.8M$ in the equatorial plane for different values of the parameters (top) $\alpha_{13}$ and (bottom) $\alpha_{22}$. In each panel, only one parameter is varied, while the other one is set to zero. This element indicates the amount of frame-dragging, which depends strongly on the deviation parameters in the vicinity of the event horizon. The vertical dashed line marks the location of the event horizon.}
\label{fig:framedragging}
\end{figure}

Accretion flows in three-dimensional fully relativistic magnetohydrodynamic simulations are greatly affected by the amount of frame-dragging near the black hole, while the location of the ISCO is only of marginal importance. In order to assess the amount of frame-dragging of the lowest-order metric, I expand the $(t,\phi)$ component of the metric in Eq.~\eqref{eq:metric_KS} in $1/r$ and obtain the expression
\ba
g_{t\phi}^{\rm KS} &=& -\frac{2aM\sin^2\theta}{r} \left[ 1 + \frac{\alpha_{22}M}{2r} \right. \nn \\
&& \left. - \frac{2a^2\cos^2\theta-\alpha_{13}M^2}{2r^2} + {\cal O}\left(\frac{1}{r^3}\right) \right].
\ea
The leading nonvanishing order of the deviations from the Kerr metric is of order $1/r^2$, while the next-order correction of the Kerr part is only of order $1/r^3$. Consequently, frame-dragging can be strongly affected by the presence of non-Kerr deviations, and positive values of the parameters $\alpha_{22}$ and $\alpha_{13}$ enhance the amount of frame-dragging. I illustrate the dependence of the $(t,\phi)$ element of the metric in Kerr-Schild-like form on the parameters $\alpha_{13}$ and $\alpha_{22}$ in Fig.~\ref{fig:framedragging}.

\section{Mapping to Other Metrics}
\label{sec:mapping}

In this section, I derive the explicit mapping of the metric given in Eq.~\eqref{eq:metric} to other known metrics from the literature. In particular, these include known four-dimensional, analytic black hole solutions in modified theories of gravity. These metrics can be written as
\be
g_{\mu\nu} \equiv g_{\mu\nu}^{\rm K} + h_{\mu\nu},
\ee 
where $h_{\mu\nu}$ is the modification of the Kerr part of the metric, which depends on one or more parameters. I will match metrics of this form with the metric in Eq.~\eqref{eq:metric}, which I expand to linear order in the deviation parameters.

\subsection{The modified gravity bumpy Kerr metric}

The metric designed in Ref.~\cite{VYS11} (the ``modified gravity bumpy Kerr metric'', hereafter labeled ``MGBK'') derives from the most general stationary, axisymmetric metric in Lewis-Papapetrou form with the additional requirement that it possesses three independent constants of motion for small deviations away from the Kerr metric, where the third, Carter-like constant is quadratic in the momentum. This metric is defined in terms of a coupled set of integro-differential equations, which (in the deformed Kerr parametrization of Ref.~\cite{VYS11}) are listed in Appendix~A. Note that this metric depends on four deviation functions as is the case of the metric in Eq.~\eqref{eq:metric}, which are denoted $\gamma_i(r)$, $i=1,3,4$, and $\Theta_3(\theta)$.

Making an ansatz for the $(t,\phi)$ component of this metric in the form of a power series in $1/\sqrt{\Sigma}$,
\be
h^{\MGBK}_{t \phi} = M \sum_{n=2}^{N} h_{t\phi,n}(\theta) \left( \frac{M^n}{\Sigma^{n/2}} \right),
\ee
Gair and Yunes \cite{GY11} solved the set of integro-differential equations and obtained the remaining metric elements:
\begin{widetext}
\ba
h^{\MGBK}_{tt} &=& -\frac{a}{M} \frac{P^{}_2}{P^{}_1} h^{\MGBK}_{t\phi} - \frac{a}{2M} \frac{\Sigma^2 \Delta}{P^{}_1} \frac{\partial h^{\MGBK}_{t\phi}}{\partial r}  + \frac{(r^2+a^2)\hat{\rho}^{2} \Delta}{P^{}_1} \gamma_{1} + \frac{2a^2r^2 \Delta \sin^2\theta}{P^{}_1} \gamma_1 - \frac{a}{M} \frac{\Delta\sin^2\theta}{\Sigma} \frac{P^{}_3}{P^{}_1} \gamma_3 + \frac{2\Delta}{\Sigma} \frac{P^{}_4}{P^{}_1} \gamma_4 \nonumber \\
&& - \frac{a^2}{2M} \frac{\Sigma \Delta^2 \sin^2\theta}{P^{}_1} \frac{d\gamma_1}{dr} - \frac{a}{2M} \frac{\Delta^2 (\hat{\Sigma}+2a^2Mr\sin^2\theta) \sin^2\theta}{P^{}_1} \frac{d \gamma_3}{dr} - \frac{a^2}{2M} \frac{\Delta^2 (\Sigma-4Mr) \sin^2\theta}{P^{}_1} \frac{d\gamma_4}{dr} \,, \nonumber \\
h^{\MGBK}_{rr} &=& - \frac{\Sigma \gamma_{1}}{\Delta}  \,, \nonumber \\
h^{\MGBK}_{\theta\theta} &=& 0, \nn \\
h^{\MGBK}_{\phi\phi} &=& -\frac{(r^2+a^2)^2}{a^2} h^{\MGBK}_{tt} + \frac{\Delta}{a^2} \Sigma \gamma_{1} - \frac{2(r^2+a^2)}{a} h^{\MGBK}_{t\phi}  - \frac{2\Delta^2 \sin^2\theta}{a} \gamma_3 + \frac{2\Delta^2}{a^2} \gamma_4,
\ea
\end{widetext}
where
\ba
\hat{\rho}^{2} &\equiv& r^{2} - a^{2} \cos^{2}{\theta},
\label{eq:rhohat} \\
\Theta_3(\theta) &=& 0, \\
\gamma_{A} &=& \sum_{n=0}^{\infty} \gamma_{A,n} \left(\frac{M}{r}\right)^{n}\,,~~~A=1,4, \\
\gamma_{3} &=& \frac{1}{r} \sum_{n=0}^{\infty} \gamma_{3,n} \left(\frac{M}{r}\right)^{n}.
\ea
The functions $P_i$, $i=1-4$, are polynomials in $r$ that can be found in Appendix~A of Ref.~\cite{VYS11} and the coefficients $h_{t\phi,n}$ are given in Ref.~\cite{GY11}. Additional simplifications~\cite{GY11} allow one to set
\be
\gamma_{1,0}=\gamma_{1,1}=\gamma_{3,0}=\gamma_{3,2}=\gamma_{4,0}=\gamma_{4,1} = 0.
\label{eq:MGBKsimp1}
\ee

Gair and Yunes \cite{GY11} gave an explicit version of this metric expanding all metric components in power series in $1/r$,
\be
h_{\mu\nu}^\MGBK = \sum_n h_{\mu\nu,n} \left( \frac{M}{r} \right)^n,
\ee
where the coefficients $h_{\mu\nu,n}$ are given in Eqs.~(27)--(30) of Ref.~\cite{GY11}. Note that while current PPN constraints were taken into account in Ref.~\cite{GY11} up to ${\cal O}(1/r)$, the $h_{tt,2}$ coefficient in Eq.~(27) of Ref.~\cite{GY11} is likewise tightly constrained by current PPN experiments (c.f., Eq.~\eqref{eq:PPN_A} and Ref.~\cite{WillLRR}) and
\be
\gamma_{1,2} = \gamma_{3,1} = \gamma_{4,2} = 0
\label{eq:MGBKsimp2}
\ee
should be chosen so that $h_{tt,2}=0$. In principle, the choice $\gamma_{1,2}=-2\gamma_{4,2}$, $\gamma_{3,1}=0$ is sufficient so that the $h_{tt,2}$ coefficient complies with the PPN constraints. This choice, however, would introduce an undesirable finetuning between the parameters $\gamma_{1,2}$ and $\gamma_{4,2}$, which was also avoided in some of the other simplifications in Ref.~\cite{GY11} as listed in Eq.~\eqref{eq:MGBKsimp1}.

With this minor adjustment, I obtain from the metric elements $h_{rr}^\MGBK$, and $h_{\theta\theta}^\MGBK$ the mapping
\ba
\alpha_{5n} &=& \gamma_{1,n},~~~n\geq2, 
\label{eq:mapalpha5MGBK} \\
\epsilon_n &=& 0,~~~n\geq3,
\label{eq:mapepMGBK} 
\ea
as well as from the elements $h_{tt}^\MGBK$, $h_{t\phi}^\MGBK$ the mapping 
\ba
\sum_{n=3}^\infty \alpha_{1n} \left(\frac{M}{r}\right)^n &=& \frac{1}{4(r^2+a^2)\Delta} \{ 8aMrh_{t\phi}^\MGBK \nn \\
&& +[2r^4+a^4 +a^2r(3r+4M) \nn \\
&& +a^2(\Delta-2Mr)\cos2\theta]h_{tt}^\MGBK  \}, 
\label{eq:mapalpha1MGBK} \\
\sum_{n=2}^\infty \alpha_{2n} \left(\frac{M}{r}\right)^n &=& -\frac{1}{2a\Delta} [a(\Sigma-4Mr)h_{tt}^\MGBK \nn \\
&& + 2(\Sigma-2Mr)\csc^2\theta h_{t\phi}^\MGBK ],
\label{eq:mapalpha2MGBK}
\ea
where the right-hand side of Eqs.~\eqref{eq:mapalpha1MGBK}--\eqref{eq:mapalpha2MGBK} has to be expanded in $1/r$. Should any terms in this expansion depend on the polar angle $\theta$, the coefficients of these terms have to be set to zero by an appropriate choice of the deviation parameters $\gamma_{k,n}$, $k=1,3,4$, $n\geq2$, because the left-hand side in these equations are functions of radius only. As I will show below, no such angular terms occur at least up to order $n=5$. Note that the remaining matching of the metric element $h_{\phi\phi}^\MGBK$ with the deviation from the $(\phi,\phi)$ element of the metric in Eq.~\eqref{eq:metric} linearized in the deviation parameters serves here as a consistency check of the mapping of the deviation parameters in Eqs.~\eqref{eq:mapalpha1MGBK}--\eqref{eq:mapalpha2MGBK}, which may impose additional requirements on certain of the parameters $\gamma_{k,n}$. This is likewise not the case at least up to ${\cal O}(1/r^5)$.

The coefficients $h_{\mu\nu,n}^\MGBK$ are given in Ref.~\cite{GY11} up to ${\cal O}(1/r^5)$, and up to this order I find for the coefficients of Eqs.~\eqref{eq:mapalpha1MGBK}--\eqref{eq:mapalpha2MGBK} the explicit relations
\ba
\alpha_{13} - \frac{\alpha_{53}}{2} &=& \gamma_{4,3}, \nn \\
\alpha_{14} - \frac{\alpha_{54}}{2} &=& - 2\gamma_{4,3} + \gamma_{4,4}, \nn \\
\alpha_{15} - \frac{\alpha_{55}}{2} &=& - 2\gamma_{4,4} + \gamma_{4,5}, \nn \\
\alpha_{22} &=& \frac{M}{a}\gamma_{3,3}, \nn \\
\alpha_{23} - \frac{\alpha_{53}}{2} &=& -\frac{M}{a}(2\gamma_{3,3}-\gamma_{3,4}), \nn \\
\alpha_{24} - \frac{\alpha_{54}}{2} &=& \frac{a}{M}\gamma_{3,3} - \frac{M}{a}(2\gamma_{3,4}-\gamma_{3,5}), \nn \\
\alpha_{25} - \frac{\alpha_{55}}{2} &=& \frac{a}{M}\gamma_{3,4} - \frac{M}{a}(2\gamma_{3,5}-\gamma_{3,6}).
\ea

These equations suggest a general mapping of the form
\ba
\alpha_{1n} - \frac{\alpha_{5n}}{2} &=& -2\gamma_{4,n-1} + \gamma_{4,n}, \nn \\
\alpha_{2n} - \frac{\alpha_{5n}}{2} &=& \frac{a}{M}\gamma_{3,n-1} - \frac{M}{a}(2\gamma_{3,n}-\gamma_{3,n+1}),
\ea
where the first equation holds for $n\geq3$, while the second equation holds for $n\geq2$ [note Eqs.~\eqref{eq:MGBKsimp1} and \eqref{eq:MGBKsimp2}]. I have not investigated this mapping at orders $n>5$. If this mapping is valid at all orders $n$, then, together with Eqs.~\eqref{eq:mapalpha5MGBK} and \eqref{eq:mapepMGBK}, the linearized form of the metric in Eq.~\eqref{eq:metric} and the metric of Ref.~\cite{VYS11} in the above form \cite{GY11} are equivalent. Eq.~\eqref{eq:mapepMGBK} reduces the number of deviation functions of the metric in Eq.~\eqref{eq:metric} for this mapping to three, which is in accordance with the choice of Gair and Yunes \cite{GY11} to set $\Theta_3(\theta)=0$. 

Due to the implicit form of some of the elements in the general metric of Ref.~\cite{VYS11} in Eq.~\eqref{eq:MGBK} I do not attempt to map it to the linearized form of the metric in Eq.~\eqref{eq:metric} or in Eq.~\eqref{eq:metric_gen}. Should it turn out that these two metrics cannot be mapped exactly, then it must be possible to further generalize the metric designed in this paper, because the metric in Eq.~\eqref{eq:MGBK} was obtained from the most general stationary, axisymmetric metric in Lewis-Papapetrou form which admits three constants of motion for small deviations from the Kerr metric and because the Carter-like constant in both metrics is quadratic in the momentum (c.f., Eq.~\eqref{eq:modCarterConst} in this paper and Eq.~(38) in Ref.~\cite{VYS11}).

\subsection{Einstein-Dilaton-Gauss-Bonnet Gravity}

Static black holes in gravity theories described by Lagrangians modified from the standard Einstein-Hilbert form by scalar fields coupled to quadratic curvature invariants were investigated in Ref.~\cite{YS11}. In these solutions, the relevant component of the metric deformation is given by
\ba
&& h_{rr}^{\rm EDGB} =  - \frac{\alpha_3}{\kappa M^2 r^2 f_S(r)^2}\nn \\
&& \left( 1 + \frac{M}{r} + \frac{52}{3}\frac{M^2}{r^2} + \frac{2M^3}{r^3} + \frac{16}{5}\frac{M^4}{r^4} - \frac{368}{3}\frac{M^5}{r^5} \right),
\ea
where $\alpha_3$ is the coupling constant of this theory and \mbox{$\kappa = 1/(16\pi)$}.

The mapping is, then, given by the equation
\ba
&& \sum_{n=2}^\infty \alpha_{5n} \left( \frac{M}{r} \right)^n = \frac{\alpha_3 M^2}{15 \kappa r^6(r-2M)} (1840M^5-48M^4r \nn \\
&& -30M^3r^2-260M^2r^3-15Mr^4-15r^5)
\ea
and the lowest-order coefficients are:
\ba
\alpha_{52} &=& -\frac{\alpha_3}{\kappa}, \nn \\
\alpha_{53} &=& -\frac{3\alpha_3}{\kappa}, \nn \\
\alpha_{54} &=& -\frac{70\alpha_3}{3\kappa}.
\ea

\subsection{Chern-Simons Gravity}

Slowly rotating black holes in dynamical Chern-Simons gravity were analyzed in Ref.~\cite{YP09}. In these solutions, only the $(t,\phi)$ component of the metric is modified, which is given by the expression
\be
h_{t\phi}^{\rm CS} = \frac{5}{8}\zeta_{\rm CS}\frac{a}{M}\frac{M^5}{r^4}\sin^2\theta \left(1+\frac{12M}{7r}+\frac{27M^2}{10r^2}\right).
\ee
In this case, the mapping is
\ba
\alpha_{24} &=& \frac{5}{8}\zeta_{\rm CS}, \\
\alpha_{25} &=& \frac{15}{14}\zeta_{\rm CS}, \\
\alpha_{26} &=& \frac{27}{16}\zeta_{\rm CS}.
\ea
All other deviation parameters vanish. Note that the metric to ${\cal O}(a^2)$ found in Ref.~\cite{YYT12} is not integrable and, thus, cannot be mapped to the metric in Eq.~\eqref{eq:metric}.

\subsection{Braneworld black holes}

One class of metrics that cannot be related to the metric in Eq.~\eqref{eq:metric} via a simple mapping is the rotating black hole solution in Randall-Sundrum-type braneworld gravity \cite{RS2}, which was found in Ref.~\cite{braneBH}. This metric is given by the elements \cite{braneBH}
\ba
g_{tt}^{\rm RS2} &=& -\left(1-\frac{2Mr-\beta}{\Sigma}\right), \nn \\
g_{rr}^{\rm RS2} &=& \frac{\Sigma}{\bar{\Delta}}, \nn \\
g_{\theta\theta}^{\rm RS2} &=& \Sigma, \nn \\
g_{\phi\phi}^{\rm RS2} &=& \left( r^2+a^2+\frac{2Mr-\beta}{\Sigma}a^2\sin^2\theta \right)\sin^2\theta, \nn \\
g_{t\phi}^{\rm RS2} &=& \frac{a(2Mr-\beta)\sin^2\theta}{\Sigma},
\label{eq:BBHmetric}
\ea
where
\be
\bar{\Delta}\equiv\Delta+\beta
\ee
and where $\beta$ is the tidal charge which can be positive or negative.

This metric is identical to the Kerr-Newman metric where the tidal charge is simply the square of the electric charge $Q_{\rm el}$. Therefore, this metric could be trivially included in the class of metrics designed in this paper by starting with the Kerr-Newman metric instead of the Kerr metric in Eq.~\eqref{eq:contraKerr} and replacing $Q_{\rm el}^2\rightarrow\beta$.


\section{Discussion}
\label{sec:discussion}

In this paper, I designed a Kerr-like black hole metric which is regular, admits three independent, exact constants of motion, and depends on four deviations functions in a nonlinear manner. This metric contains the Kerr metric as the special case when all deviations vanish. This metric does not derive from any particular set of field equations, but, instead, can be used as a framework for model-independent strong-field tests of the no-hair theorem with observations of black holes in the electromagnetic spectrum. I showed that the event horizon of this metric is identical to the event horizon of the Kerr metric in Boyer-Lindquist coordinates and that the Killing horizon coincides with the event horizon. I also determined the range of the deviation parameters for which the event horizon and the exterior domain are free of pathologies such as singularities or closed timelike curves.

When expanded to linear order in the deviation parameters, this metric can be related to the Kerr-like metric of Ref.~\cite{VYS11}. I found an explicit general mapping between the new metric and the metric of Ref.~\cite{VYS11} in the form found in Ref.~\cite{GY11}, which holds at least up to the fifth order in an expansion of the metric elements in $1/r$. I likewise mapped my new metric to known four-dimensional, analytic black hole solutions in modified theories of gravity.

I showed that the equations of motion for a test particle on a geodesic orbit in this metric can be written in first-order form and found the Carter-like constant which is the third constant of motion along the orbit of the particle in addition to its energy and axial angular momentum. For particles on circular equatorial orbits, I derived expressions for its energy, axial angular momentum, and dynamical frequencies and I calculated the location of the ISCO. I showed that these quantities depend significantly on the deviation parameters.

These properties make this metric a well-suited framework for strong-field tests of the no-hair theorem in the electromagnetic spectrum. Thanks to the existence of an event horizon and the regularity of the exterior domain, both geometrically thin and thick accretion flows can be properly modeled (see the discussion in Refs.~\cite{PaperI,Joh13}). The first-order form of the equations of motion increases the speed and precision of ray-tracing codes that are used to model and predict observational signatures of non-Kerr black holes, because these otherwise have to solve the second-order geodesic equations (see Refs.~\cite{PaperI,PJ11}). In addition, for tests of the no-hair theorem in the gravitational-wave spectrum, the mapping of the metric linearized in the deviations parameters establishes the connection to the metric of Ref.~\cite{VYS11}, for which approximate EMRI were constructed in Ref.~\cite{GY11}.

Potential deviations from the Kerr metric should be observable with several techniques in the electromagnetic spectrum including the continuum-fitting and iron line methods. See Refs.~\cite{CFironReviews} for comprehensive reviews on these methods. These two methods directly measure the location of the ISCO. Since the ISCO does not depend on the deviation function $A_5(r)$, this type of deviation would be difficult to detect with these techniques. Such particular deviations, however, may be detected with very-long baseline imaging observations of supermassive black holes (c.f., e.g., Ref.~\cite{PaperII}) or of quasi-periodic variability (c.f., e.g., Ref.~\cite{PaperIII}).

In order to facilitate fully relativistic magnetohydrodynamic simulations of accretion flows in the new metric, I constructed a transformation to Kerr-Schild-like coordinates, which properly removes the coordinate singularity at the event horizon in Boyer-Lindquist-like coordinates. I demonstrated that the lapse is positive across the event horizon and that the amount of frame-dragging near the black hole depends strongly on the deviation parameters, which can greatly amplify the amount of frame-dragging induced by the spin alone. Such accretion flow simulations are carried out in a stationary black hole background where the dynamical properties of the gravity theory do not need to be known. Instead, only the coupling of matter to electromagnetic fields in the modified theory needs to be specified. A first step is to assume that such interactions are governed by the laws of Maxwell electrodynamics.

Refs.~\cite{GMcKT03,McKG04} implemented a general-relativistic magnetohydrodynamic code that is based on a stationary black hole metric and several technical assumptions which, among others, allow for the treatment of the plasma as a perfect fluid. These assumptions also require that the exterior domain endowed with the metric is globally hyperbolic. While this property is not proven here, it seems plausible that global hyperbolicity holds in the Kerr-like metric designed in this paper, because the exterior domain is regular, and, thus, it should be possible to trace any flow particle outside of the event horizon uniquely to future infinity (or the horizon) for any set of initial conditions. Simulations of this nature should, therefore, also be feasible in this metric and will be explored in detail in a future paper.

\acknowledgments
I thank A. Broderick and J. McKinney for conversations on magnetohydrodynamic simulations. This work was supported by a CITA National Fellowship at the University of Waterloo and in part by Perimeter Institute for Theoretical Physics. Research at Perimeter Institute is supported by the Government of Canada through Industry Canada and by the Province of Ontario through the Ministry of Research and Innovation.


\appendix
\section{General Form of the MGBK Metric}

Here I write explicitly the general form of the metric of Ref.~\cite{VYS11}, where a slightly different notation was used. In the notation I use in this paper, the metric is given by the elements
\begin{widetext}
\ba
h^{\MGBK}_{tt} &=& -\frac{a}{M} \frac{P_2}{P_1} h^{\MGBK}_{t\phi} - \frac{a}{2M} \frac{\Sigma^2 \Delta}{P_1} \frac{\partial h^{\MGBK}_{t\phi}}{\partial r} - \frac{2a^2r (r^2+a^2) \Delta \sin\theta \cos\theta}{\Sigma P_1} h^{\MGBK}_{r\theta} + \frac{(r^2+a^2)\hat{\rho}^{2} \Delta}{\Sigma P_1} \I \nonumber \\
&& + \frac{2a^2r^2 \Delta \sin^2\theta}{P_1} \gamma_1 + \frac{\hat{\rho}^{2} (r^2+a^2) \Delta}{\Sigma P_1} \Theta_3 - \frac{a}{M} \frac{\Delta\sin^2\theta}{\Sigma} \frac{P_3}{P_1} \gamma_3 + \frac{2\Delta}{\Sigma} \frac{P_4}{P_1} \gamma_4 \nonumber \\
&& - \frac{a^2}{2M} \frac{\Sigma \Delta^2 \sin^2\theta}{P_1} \frac{d\gamma_1}{dr} - \frac{a}{2M} \frac{\Delta^2 (\Sigma+2a^2Mr\sin^2\theta) \sin^2\theta}{P_1} \frac{d \gamma_3}{dr} - \frac{a^2}{2M} \frac{\Delta^2 (\Sigma-4Mr) \sin^2\theta}{P_1} \frac{d\gamma_4}{dr} \,, \nonumber \\
h^{\MGBK}_{rr} &=& - \frac{1}{\Delta} \I -\frac{1}{\Delta} \Theta_3 \,,
\nonumber \\
h^{\MGBK}_{\phi\phi} &=& -\frac{(r^2+a^2)^2}{a^2} h^{\MGBK}_{tt} + \frac{\Delta}{a^2} \I - \frac{2(r^2+a^2)}{a} h^{\MGBK}_{t\phi} + \frac{\Delta}{a^2} \Theta_3 - \frac{2\Delta^2 \sin^2\theta}{a} \gamma_3 + \frac{2\Delta^2}{a^2} \gamma_4 \,, \nonumber \\
\frac{\partial h^{\MGBK}_{\theta\theta}}{\partial r} &=& \frac{2r}{\Sigma} h^{\MGBK}_{\theta\theta} + \frac{2a^2 \sin\theta \cos\theta}{\Sigma} h^{\MGBK}_{r\theta} + 2 \frac{\partial h^{\MGBK}_{r\theta}}{\partial \theta} + \frac{2r}{\Sigma} \I - 2r \, \gamma_1 + \frac{2r}{\Sigma} \Theta_3 \,, \nonumber \\
\frac{\partial^2 h^{\MGBK}_{t\phi}}{\partial r^2} &=& \frac{8aM \sin\theta \cos\theta}{\Sigma^4} \frac{P_5}{P_1} h^{\MGBK}_{r\theta} - \frac{4aMr(r^2+a^2) \sin\theta \cos\theta}{\Sigma^3} \frac{\partial h^{\MGBK}_{r\theta}}{\partial r} + \frac{2a^2\sin^2\theta}{\Sigma^2} \frac{P_6}{P_1} h^{\MGBK}_{t\phi} \nn \\
&& - \frac{2r}{\Sigma}\frac{P_7}{P_1} h^{\MGBK}_{t\phi} + \frac{4aMr\sin^2\theta}{\Sigma^2}\frac{P_{15}}{P_{16}} \I - \frac{4aMr\sin^2\theta}{\Sigma^2}\frac{P_8}{P_1} \gamma_1 + \frac{4aMr}{\Sigma^2}\frac{P_9}{P_1} \Theta_3 \nn \\
&& + \frac{2\sin^2\theta}{\Sigma^2}\frac{P_{10}}{P_1} \gamma_3 - \frac{16aM \sin^2\theta}{\Sigma^2}\frac{P_{11}}{P_1} \gamma_4 - \frac{2a}{\Sigma^2}\frac{P_{12}}{P_1} \frac{d\gamma_1}{dr} - \frac{2\sin^2\theta}{\Sigma^2}\frac{P_{13}}{P_1} \frac{d\gamma_3}{dr} \nn \\
&& - \frac{2a\sin^2\theta}{\Sigma^2}\frac{P_{14}}{P_1} \frac{d\gamma_4}{dr} - \frac{a \Delta \sin^2\theta}{\Sigma} \frac{d^{2} \gamma_1}{dr^{2}} - \frac{\Delta \sin^2\theta}{\Sigma^2}(\Sigma+2a^2Mr\sin^2\theta) \frac{d^{2} \gamma_3}{dr^{2}} \nn \\
&& - \frac{a \Delta (\Sigma-4Mr) \sin^2\theta}{\Sigma^2} \frac{d^{2} \gamma_4}{dr^{2}} \,,
\label{eq:MGBK}
\ea
\end{widetext}
where
\ba
\Theta_3&=&\Theta_3(\theta), \\
\gamma_i&\equiv&\gamma_i(r),~~~i=1,3,4,
\ea
are arbitrary functions of the polar angle $\theta$ and the radius $r$, respectively, and
\be
\I \equiv \int dr \left[ \frac{2a^2 \sin\theta \cos\theta}{\Sigma} h^{\MGBK}_{r\theta} + 2r \, \gamma_1 + \Sigma \, \frac{d \gamma_1}{dr} \right].
\ee
The function $\hat{\rho}$ is given in Eq.~\eqref{eq:rhohat} and the functions $P_j$, $j=1-15$, are polynomials in $r$ and $\cos\theta$, given explicitly in Appendix~A of Ref.~\cite{VYS11}.

\end{document}